\documentclass[aps,pre,showpacs,preprintnumbers,amsmath,amssymb]{revtex4}
\usepackage{natbib}
\usepackage{amsfonts,amssymb,amsmath,amsbsy,amscd}
\usepackage{latexsym,euscript,exscale,epic,eepic,epsfig}
\usepackage{subfigure}
\usepackage{times}

\begin{document}
\title{Localization of elastic deformation in strongly anisotropic,
porous, linear materials with periodic microstructures: exact solutions and
dilute expansions}

\author{Francois Willot}
\email{francois.willot@polytechnique.edu}
\altaffiliation{D\'epartement de M\'ecanique, \'Ecole Polytechnique, 91128
 Palaiseau Cedex, France.}
\altaffiliation{Department of Mechanical Engineering and
Applied Mechanics, University of Pennsylvania, Philadelphia PA 19104-6315 USA.}
\author{Yves-Patrick Pellegrini}
\affiliation{D\'epartement de
Physique Th\'eorique et Appliqu\'ee, Commissariat \`a l'\'Energie
Atomique, BP12, 91680 Bruy\`eres-le-Ch\^atel, France.}
\email{yves-patrick.pellegrini@cea.fr}
\author{Pedro Ponte Casta\~neda}
\email{ponte@seas.upenn.edu}
\affiliation{D\'epartement de M\'ecanique, \'Ecole Polytechnique, 91128
 Palaiseau Cedex, France.}
\affiliation{Department of Mechanical Engineering and
Applied Mechanics, University of Pennsylvania, Philadelphia PA 19104-6315 USA.}

\pacs{A voids and inclusions; B anisotropic material; B constitutive
behaviour; C Energy methods; localization.}

\begin{abstract}
Exact solutions are derived for the problem of a two-dimensional, infinitely
ani\-so\-tro\-pic, linear-elastic medium containing a periodic lattice of
voids. The matrix material possesses either one infinitely soft, or one
infinitely hard loading direction, which induces localized (singular) field
configurations. The effective elastic moduli are computed as functions of the
porosity in each case. Their dilute expansions feature half-integer powers of
the porosity, which can be correlated to the localized field patterns.
Statistical characterizations of the fields, such as their first moments and
their histograms are provided, with particular emphasis on the singularities of
the latter. The behavior of the system near the void close packing fraction is
also investigated. The results of this work shed light on corresponding results
for strongly nonlinear porous media, which have been obtained recently by means
of the ``second-order'' homogenization method, and where the dilute estimates
also exhibit fractional powers of the porosity.
\end{abstract}

\maketitle

\section{Introduction}
\subsection{Context}
Most of the currently available homogenization methods for strongly non-linear
composites make use an underlying linear homogenization estimate, whether aimed
at computing dielectric (Willis, 1986; Zeng et al., 1988; Ponte
Cas\-ta\-\~ne\-da, De\-Bot\-ton, Li, 1992), or elastic-plastic transport
properties (Ponte Cas\-ta\-\~ne\-da, 1991; Pon\-te Cas\-ta\-\~ne\-da, 1996;
Mas\-son et al., 2000; Ponte Cas\-ta\-\~ne\-da, 2002; La\-hel\-lec and
Su\-quet, 2004). The best results are obtained using an
\emph{an\-iso\-tro\-pic} linear estimate, the an\-iso\-tro\-py of which is
consistently determined, by the means and variances of the fields in each phase
of the composite (Pellegrini, 2001; Ponte Casta\~neda, 2001). The
an\-iso\-tro\-py of the underlying linear medium accounts for the privileged
direction imposed by the driving field in the non-linear medium.

Applying the ``second-order'' method (Ponte Casta\~neda, 2002) to a random
power-law material weakened by aligned cylindrical voids, it has been found
that in the so-called ``dilute" limit (i.e.,\ that of a vanishingly small
volume fraction of porosity $f$), and in the limit of infinite exponent where
the matrix becomes ideally plastic with a yield threshold, the leading
correction to the yield stress in pure shear behaves as $f^{2/3}$. This result
holds for both Hashin-Shtrikman and self-consistent estimates. The second-order
method of Pellegrini (2001), suitably modified by replacing a Gaussian ansatz
for the field distributions by a Heaviside ansatz (so as to cut-off high fields
in a way compatible with threshold-type materials), provides in some cases
similar predictions (Pellegrini and Ponte Casta\~neda, 2001, unpublished). This
$f^{2/3}$ dilute behavior is unusual in the context of effective-medium
theories for \emph{random} media, where the low-order terms in dilute
expansions from the literature usually consist in integer powers of $f$. With
the $f^{2/3}$ correction, the derivative of the yield stress with respect to
the porosity is infinite at $f=0$, due to the exponent being lower than 1.
Thus, a vanishingly small volume fraction of voids induces a dramatic weakening
of the porous medium. Ponte Casta\~neda (1996, 2002) interprets this phenomenon
as an indication of localizing behavior in a regime where shear bands pass
through the pores, remarking that the limit analysis of Drucker (1966) produces
an upper bound for the yield threshold with a dilute correction behaving as
$f^{1/2}$ in 2D. Numerical calculations and tests on perforated plates with
periodically distributed holes (Francescato and Pastor, 1998) are consistent
with these predictions in a plane stress situation. However, for a square
network of circular holes, kinematic and static limit analyses result in the
following analytical bounds on the effective yield stress $\widetilde{Y}$
(Francescato et al., 2004):
\begin{equation}
\label{bounds}
1-2(f/\pi)^{1/2}\leq \widetilde{Y}/Y\leq (2/\sqrt{3})\left[1-2(f/\pi)^{1/2}\right],
\end{equation}
where $Y$ is the yield threshold of the matrix. Numerical limit-analysis on a
hollow disk model in plane strain leads to somewhat different conclusions, with
a correction behaving roughly as $f^{2/3}$ for uniform strain boundary
conditions,and as $f^{1/2}$ for uniform stress boundary conditions (Pastor and
Ponte Casta\~neda,2002). However, the calculations were not sufficiently
accurate to be completely definitive. On the other hand, the exponent is quite
certainly associated with localizing behavior, and its presence seems to be
independent on whether the system is random, periodic, or comprises a unique
void (although the actual values of the exponent may be dependent on the
specific microstructure considered).

Because the second-order theory with exponent $2/3$ inherits its properties
from an underlying anisotropic linear effective-medium theory, whose anisotropy
is consistently determined by the field variances, a natural question concerns
the ``localizing'' properties of the anisotropic linear theory itself, and its
relation with possible non-analyticities of the effective shear moduli in the
dilute limit. Bearing these considerations in mind, the present paper focuses
on a new \emph{exact} solution for the special, but representative, case of a
two-dimensional (2D) array of voids embedded in an anisotropic linear elastic
matrix, in the singular limit of infinite anisotropy. We emphasize that no
exact solution of a similar type is available for general non-linear periodic
media, which provides additional motivation for undertaking this study.

\subsection{Organization of the paper}
The constitutive laws, the prescribed loading conditions and the notations are
defined in Sec.\ \ref{sec:setup}. The word ``loading'' refers to the
prescription of either overall conditions of homogeneous stress, or of
homogeneous strain in the linear medium. Both are equivalent since they are
related by the effective moduli. The matrix in the composite is compressible
and possesses anisotropic properties in shear. Two special limits of infinite
anisotropy which lead to exact results are then considered (Secs.\
\ref{sec:ellzero} and \ref{sec:ellinfty}). For each of these limits, simple
shear, pure shear, and equibiaxial loading situations are examined, and
solutions are provided in each case (for some loading and anisotropy conditions
however, only the limit of an incompressible matrix is considered; but this
limit is the one relevant to non-linear homogenization of plastic porous
media). The effective moduli and their dilute expansions, together with the
mean and variance of the strain and stress fields are computed in each case.
All these quantities are relevant to non-linear homogenization theories. A
useful way of condensing the information contained in the solutions is by using
field distributions which can in principle directly be used to compute the
effective energy (Pellegrini, 2001) or, more trivially, the moments of the
fields. Local extrema or saddle points in the field maps induce singularities
in the distributions such as power or logarithmic divergences, or
discontinuities, of the types observed in the density of states of a crystal
(Van Hove 1953; Abrikosov, Campuzano and Gofron, 1993). These singularities are
examined in the context of our exact results. Cule and Torquato (1998) computed
analytically the distributions of the electric field in a particular
dielectric/conductor composite. They found singularities of the Van Hove type,
but no extended singularities of the type put forward by Abrikosov \textit{et
al.} We extract below some singularities of this type, while singularities of
yet a different type are encountered in Sec.\ \ref{sec:vh2}. Obviously, in the
periodic case, the field distributions are redundant with the exact solutions
derived hereafter. However, for randomly disordered situations where exact
solutions are not available, they remain the only way to collect useful
informations on the fields. To ease their interpretation, a knowledge of their
features in the periodic case is desirable, which provides a justification for
the studies of Secs.\ \ref{sec:vh1} and \ref{sec:vh2}. Moreover, we find that
the infinite field variances obtained for some particular loadings are
correlated to field singularities directly linked to localization patterns,
which have counterparts as characteristic features in the distributions. A
summary of our findings and a conclusion close the paper (Sec.\
\ref{sec:concl}).

\section{Problem formulation}
\label{sec:setup}
\subsection{Material constitutive law and microstructure}
\begin{figure}[h!tbp]
\begin{center}
\epsfig{file=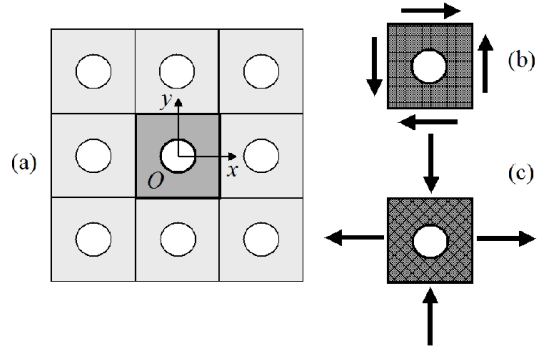,width=7cm}
\begin{minipage}{12cm}
\caption{\label{fig:microstructure} Left, periodic porous medium
with unit cell and reference axes (a). Right, unit cell with (b)
0-type fibers; and (c) 45-type fibers. The black arrows depict
eigenmodes of strain: simple shear (SS) in (b), and pure shear (PS)
in (c).}
\end{minipage}
\end{center}
\end{figure}

We consider a periodic porous medium, with a square unit cell of size $L$ made
of a linear-elastic matrix (denoted by a phase index $\beta=1$), containing a
single disk-shaped void of radius $a$ (phase index $\beta=2$). Cartesian
reference axes $Ox$, $Oy$ are defined as in Fig.\ \ref{fig:microstructure}a,
such that the void center lies at $(x,y)=(0,0)$. The unit cell is the square
region $[-L/2, L/2]\times[-L/2, L/2]$. The porosity is the surface
concentration of voids $f=\pi(a/L)^2$. The close-packing value where the voids
touch is $f_c=\pi/4\simeq 0.78$, for $a=L/2$. Hereafter, all lengths are
rescaled by $L$, so that $L\equiv 1$. The following notations are employed
throughout the text: the \emph{upper-right quadrant} (URQ) of the unit cell
denotes the region $(x,y)\in[0,1/2]^2$, the \emph{lower-right quadrant} (LRQ)
stands for $(x,y)\in[0,1/2]\times[-1/2,0]$. Similar abbreviations ULQ and LLQ
stand for the corresponding left quadrants.
Small deformations and plane strain conditions are assumed, so that the strain
$\mbox{\boldmath$\varepsilon$}$ derives from the two-dimensional (in-plane)
displacement field $\mathbf{u}$, and $\varepsilon_{ij}=0$ if $i$ or $j=z$.
In-plane stress equilibrium requires $\partial_i \sigma_{ij}=0$, $i,j=1,2$ and
we take the constitutive relation such that $\sigma_{xz}=\sigma_{yz}=0$, so
that the problem is two-dimensional. The linear constitutive relation in the
medium is $ \mbox{\boldmath $\sigma$}
(\mathbf{x})=\mathbb{L}(\mathbf{x}):\mbox{\boldmath
$\varepsilon$}(\mathbf{x})$. In the voids, $\mathbb{L}(\mathbf{x})\equiv 0$. In
the anisotropic matrix phase, $\mathbb{L}(\mathbf{x})$ has components (Latin
indices henceforth vary over 1
 and 2):
\begin{equation}
\label{eq:Lmat} L_{ij,kl}^{(1)}=2 \kappa\, J_{ij,kl}+2\lambda\,
E^{\rm SS}_{ij,kl}+2\mu\, E^{\rm PS}_{ij,kl},
\end{equation}
where $\kappa$ is the bulk compressibility modulus, where $\lambda$, $\mu$ are
in-plane an\-iso\-tro\-pic shear moduli, and where the operators $\mathbb{J}$,
$\mathbb{E}^{\rm SS}$, $\mathbb{E}^{\rm PS}$ are mutually orthogonal projectors
of components (no summation over repeated indices here):
\begin{equation}
J_{ij,kl}=\frac{1}{2}\delta_{ij}\,\delta_{kl},\quad E^{\rm
SS}_{ij,kl}=\frac{1}{2}(1-\delta_{ij})(1-\delta_{kl}),\quad E^{\rm
PS}_{ij,kl}= \delta_{ij}\,\delta_{kl}(\delta_{ik}-1/2).
\end{equation}
A symmetric two-dimensional tensor $\mathsf{a}$ is expanded as
$\mathsf{a}=a_\text{m}\,\mathsf{I}+a_{\text{SS}}\,\mathsf{e}_{SS}+a_{\rm
PS}\,\mathsf{e}_{\rm PS}$ where $\mathsf{I}$ is the $2\times 2$ identity matrix
and where:
\begin{equation}
\mathsf{e}_{\rm SS}=\left(
 \begin{array}{cc}
 0& 1\\
 1 & 0
 \end{array}
\right) ,\qquad
 \mathsf{e}_{\rm PS}=\left(
\begin{array}{cc}
 1& 0\\
 0 & -1
 \end{array}\right),
\end{equation}
the eigenvector sets of which are related by a 45-degree rotation.
Corresponding strain modes are represented by arrows in Fig.\
\ref{fig:microstructure}b and \ref{fig:microstructure}c. The \emph{equibiaxial}
component of $\mathsf{a}$ is $a_m\equiv(a_{xx}+a_{yy})/2$. We call the second
and third components the \emph{simple shear} (SS) and \emph{pure shear} (PS)
components, with $a_\text{SS}\equiv a_{xy}$ and $a_\text{\rm
PS}\equiv(a_{xx}-a_{yy})/2$. Then $ \mathbb{J}:\mathsf{a}=a_m\mathsf{I}$, $
\mathbb{E}^{\rm SS}:\mathsf{a}=a_{\text{SS}}\,\mathsf{e}_{SS}$ and
$\mathbb{E}^{\rm PS}:\mathsf{a}=a_{\text{PS}}\,\mathsf{e}_{\rm PS}$.
Accordingly, the constitutive relations decompose into:
\begin{equation}
\label{eq:lambdamu} \sigma_{\rm SS}=2\lambda\,\varepsilon_{\rm SS},\qquad
\sigma_{\rm PS}=2\mu\varepsilon_{\rm PS}, \qquad
\sigma_\text{m}=2\kappa\,\varepsilon_\text{m}.
\end{equation}

Elastic tensor (\ref{eq:Lmat}) is of a special type of orthotropy in the plane
with $L_{1111} =L_{2222}\not=L_{1122}\not=L_{1212}$. When $\lambda >\mu$, this
type of anisotropic response could correspond to a fiber-reinforced material
with fibers aligned with the cell axes (see Fig.\ \ref{fig:microstructure}b).
On the other hand, when $\mu>\lambda$, it would correspond to a material with
reinforcing fibers aligned with the $\pm 45^{\rm o}$ direction (see Fig.\
\ref{fig:microstructure}c). We introduce a shear anisotropy ratio, $\alpha$,
and normalized shear elastic moduli, $m$ and $\ell$:
\begin{equation}
\label{eq:lm} \alpha=\lambda/\mu,\qquad m=\mu/\kappa,\qquad
\ell=\lambda/\kappa.
\end{equation}

The principal directions of the matrix being ``aligned'' with dense lines of
voids, a reinforcement of anisotropy-induced effects is expected. Besides, the
loading modes considered below are aligned with the eigenstrains. These
choices, motivated by future applications of this work to non-linear
homogenization, focus on situations most relevant to plastic localization in
porous media. Indeed, in the nonlinear theory, the loading determines the
anisotropy of the background linear medium and always coincides with one of its
eigendirections. Moreover, shear bands in random porous media quite generally
link neighboring voids together.

\subsection{Overall behavior}
In this work, we are concerned with the problem of determining the overall
behavior of the periodic, porous material described in the previous subsection.
This overall behavior is defined as the relation between the average stress
$\langle\mbox{\boldmath $\sigma$}\rangle$ and the average strain
$\langle\mbox{\boldmath $\varepsilon$}\rangle$. Under the assumption of
separation of length scales, the overall behavior of the porous material may be
determined from the effective strain potential (see Torquato, 2002)
\begin{equation}
     W = (1-f) \mathop {\min}\limits_{\mbox{\boldmath $\varepsilon$} \in \mathcal{K}} \left\langle
     w^{(1)}(\mbox{\boldmath $\varepsilon$}) \right\rangle_{(1)},
    \label{EffW}
\end{equation}
where $\left\langle \cdot \right\rangle_{(1)}$ denotes an average over the
matrix phase, $w^{(1)}(\mbox{\boldmath $\varepsilon$})=(1/2)\mbox{\boldmath
$\varepsilon$}:\mathbb{L}^{(1)}:\mbox{\boldmath $\varepsilon$}$ is the strain
potential in the matrix and $\mathcal{K}=\{ \mbox{\boldmath $\varepsilon$}\, |
\mbox{\boldmath $\varepsilon$} = (1/2)\left[\nabla \mathbf{u}+ (\nabla
\mathbf{u})^T \right],$ $\mathbf{u}=\langle\mbox{\boldmath
$\varepsilon$}\rangle\mathbf{x}+\mathbf{u}^{*},$ $\mathbf{u}^{*} \ {\rm
periodic} \}$ is the set of kinematically admissible strain fields. The overall
constitutive relation is then given by
\begin{equation}
\langle\mbox{\boldmath
$\sigma$}\rangle =\frac {\partial  W } {\partial \langle\mbox{\boldmath $\varepsilon$}\rangle}.
    \label{EffConst}
\end{equation}
Because of linearity, we define the overall elasticity tensor
$\widetilde{\mathbb{L}}$ of the porous material via the relation
$\langle\mbox{\boldmath
$\sigma$}\rangle=\widetilde{\mathbb{L}}:\langle\mbox{\boldmath
$\varepsilon$}\rangle $. This tensor can be shown to take on the same form as
(\ref{eq:Lmat}) with components ${\widetilde{L}}_{ij,kl}$ defined by effective
moduli $\widetilde{\lambda}$, $\widetilde{\mu}$ and $\widetilde{\kappa}$.

In the analyses below, it will sometimes be more convenient to work with the
\textit{effective stress potential} $U$, such that the overall constitutive
relation may be equivalently written
\begin{equation}
\langle\mbox{\boldmath $\varepsilon$}\rangle= \frac {\partial  U } {\partial \langle\mbox{\boldmath $\sigma$}\rangle},
\quad U  = (1-f) \mathop {\min}\limits_{\mbox{\boldmath
$\sigma$} \in \mathcal{S}} \left\langle u^{(1)}(\mbox{\boldmath
$\sigma$}) \right\rangle_{(1)},
 \label{EffConst_u}
\end{equation}
where $u^{(1)}(\mbox{\boldmath $\sigma$})=(1/2)\mbox{\boldmath
$\sigma$}:\left(\mathbb{L}^{(1)}\right)^{-1}:\mbox{\boldmath $\sigma$}$ is the
stress potential in the matrix and $\mathcal{S}$ denotes the set of periodic
stresses that are divergence-free in the unit cell, with prescribed average
$\langle \mbox{\boldmath $\sigma$} \rangle$, and traction-free on the
boundaries of the pores. It should be noted that the case of an isotropic
matrix ($\alpha=1$) has been addressed by McPhedran and Movchan (1994) in the
more general framework of arbitrary contrast between the matrix and the square
array of isotropic inclusions.

\subsection{Limits of infinite anisotropy and loading modes}
\label{sec:ohdc} Hereafter only the strong anisotropy limits $\alpha=0$ and
$\alpha=+\infty$, amenable to an exact solution, are considered. Therefore: (i)
When $\alpha=0$ (i.e., $\lambda=0$ or $\mu=\infty$) the medium is soft for SS
loading, and resists PS loading; (ii) When $\alpha=\infty$ (i.e.,
$\lambda=\infty$ or $\mu=0$) the medium is soft for PS loading, and resists SS
loading. Equibiaxial, SS and PS loading modes will be considered separately
hereafter, so that only one of the three averaged components of the strain or
stress is non-zero at a time. It is denoted by $\overline{\varepsilon}$ (resp.
$\overline{\sigma}$). Depending on the loading mode considered, strain and
stress components and displacements enjoy various symmetry properties
summarized in Appendix \ref{sec:sym}.

For clarity, we detail the correspondence between the loading modes and the
anisotropy properties of the material. Consider e.g. the limiting case
$\alpha=\lambda/\mu\to 0$, attained by two different types of materials: (i)
$\lambda\to 0$ and $\mu > 0$ (Material 1); (ii) $\mu\to\infty$ and
$\lambda<\infty$ (Material 2). We seek solutions with strain and stress finite
almost everywhere (a.e.), i.e., except on points or on lines in the plane. In
particular, $\sigma_\text{PS}=2\mu\,\varepsilon_\text{PS}$ is finite a.e., so
that $\varepsilon_\text{PS}\equiv 0$ a.e.\ in Material 2. Likewise
$\varepsilon_\text{SS}=\sigma_\text{SS}/(2\lambda)$
 is finite a.e., so that
 $\sigma_\text{SS}\equiv 0$ a.e.\ in Material 1.
 Hence Material 1
is infinitely soft in the SS direction and can be finitely loaded in the PS
direction, whereas Material 2 is infinitely rigid in the PS direction and can
be finitely loaded only in the SS direction. Similar considerations apply for
$\alpha=\infty$, leading to the following correspondence:
\begin{center}
 $\alpha=0$:\,\,\,\qquad PS
loading $\leftrightarrow$ $\lambda=0$,\qquad \,\,SS loading
$\leftrightarrow$ $\mu=\infty$,\\
 $\alpha=\infty$:\qquad SS loading $\leftrightarrow$
$\mu=0$, \qquad PS loading $\leftrightarrow$ $\lambda=\infty$.
\end{center}
In equibiaxial loading, both types of materials should be considered for each
of the limits  $\alpha=0$ or $\alpha=\infty$.

As is well-known, the elasticity tensor (\ref{eq:Lmat}) is positive definite
when $\lambda$, $\mu$ and $\kappa$ are all strictly positive, and we have
existence and uniqueness of solutions. As mentioned above, however, the
interest in this work is for the limiting cases where one of the eigenvalues of
the elasticity tensor (\ref{eq:Lmat}) (or, of its inverse, the compliance
tensor) tend to zero. For these cases, care must be exercised when interpreting
the solutions. As will be seen below, the relevant potential energy, or
complementary energy functionals can still be shown to be strictly convex in
the subspace of allowable strains, or stresses, and the standard theorems (see,
for example, Proposition 1.2 of Ekeland and Temam, 1974) would still ensure
existence and uniqueness of the solutions. However, appropriate components of
the strain (or stress) field can develop discontinuities in these limiting
cases. In this connection, it is also relevant to note that the governing
equations lose ellipticity, leading to hyperbolic behavior (see below). This
phenomenon is well-known in two-dimensional problems for ideally plastic
materials (Kachanov, 1974), and has been exploited in a recent study of
shape-memory polycrystals (Chenchiah and Bhattacharya, 2005). As was shown in
this later reference, the hyperbolicity of the equations can be very helpful in
interpreting the solutions obtained, and such connections will be made for the
specific cases to be considered below. In any case, it should be kept in mind
that the solutions of interest here are limiting cases of standard elasticity
problems (with positive-definite elasticity tensors) for which solutions have
been obtained by numerical methods (Willot et al., 2007).

\subsection{Hyperbolicity conditions and characteristics}
\label{sec:characteristics} Making use of the constitutive relations
(\ref{eq:lambdamu}), the stress equilibrium conditions may be written in terms
of the displacement field as a closed system of two second-order partial
differential equations (PDEs):
\begin{equation}
\left(\begin{array}{cc} \mu+\kappa & 0 \\ 0 & \lambda \end{array}\right)
 \partial_x^2 \mathbf{u}+
\left(\begin{array}{cc} \lambda & 0 \\ 0 & \mu+\kappa \end{array}\right)
 \partial_y^2 \mathbf{u}+
(\lambda+\kappa-\mu)\left(\begin{array}{cc} 0 & 1 \\ 1 & 0
\end{array}\right)
\partial_x\partial_y \mathbf{u}
=0.
\end{equation}
Following Otto et al.~(2003), this system is decoupled as:
\begin{equation}
\label{eq:Dop}
Du_i=0,\quad D \varepsilon_{ij}=0, \quad D \sigma_{ij}=0, \qquad  i=x, y, \quad j = x,y,
\end{equation}
where the PDEs for the strain and stress fields follow immediately, and where
$D$ is the fourth-order differential operator:
\begin{equation}
D = \partial_x^4+\partial_y^4
+2r\partial_x^2\partial_y^2,\quad
r=\frac{\lambda(\mu-\kappa)+2\kappa\mu}{\lambda(\mu+\kappa)}.
\end{equation}
This PDE has the symbolic form $\phi(x,y) = x^4+y^4+2rx^2y^2$, with
characteristic lines (e.g., Zachmanoglou and Thoe, 1986) defined by the
equation $\phi(\textnormal{d}x,\textnormal{d}y) = 0$, such that:
\begin{equation}
\frac{\textnormal{d}x}{\textnormal{d}y}=
\pm\sqrt{-r\pm(r^2-1)^{1/2}}.
\end{equation}
Taking $\lambda$, $\mu$, $\kappa\geq 0$, $r$ lies on the segment line
$r\in[-1,\infty)$. At any interior point $-1<r<\infty$, the quantity
$\textnormal{d}x/\textnormal{d}y$ is strictly complex; there are no
characteristic lines in the real plane and the problem is elliptic. This
corresponds to either (i) $0<\lambda,\mu<\infty$; (ii) $0<\mu<\infty$,
$\lambda=\infty$, $\kappa<\infty$; (iii) $0<\lambda<\infty$, $\mu=\infty$,
$\kappa<\infty$. At the end points $r=-1$ and $r=\infty$, the form
$\phi(\textnormal{d}x,\textnormal{d}y)$ has two distinct real roots for
$\textnormal{d}x/\textnormal{d}y$, each of them of multiplicity two. Thus the
problem becomes hyperbolic with straight characteristics $x/y=cst$:
\begin{eqnarray}
\label{eq:charact0}
\textnormal{Case 1: }&&r=\infty\quad\Leftrightarrow\quad
 \lambda=0\textnormal{ or }\mu=\kappa=\infty
 \quad\Leftrightarrow\quad x, y=cst, \\
\textnormal{Case 2: }&&r=-1\quad\Leftrightarrow\quad
 \mu=0\textnormal{ or }\lambda=\kappa=\infty
 \quad\Leftrightarrow\quad x=\pm y+cst.
\end{eqnarray}
In this study, only hyperbolic problems are considered, i.e. the medium will be
taken incompressible ($\kappa=\infty$) when $\lambda$ or $\mu=\infty$. All
characteristics encountered are straight lines aligned with either one of the
Cartesian axes ($\alpha=0$) or with one of the two diagonals of the unit cell
($\alpha=\infty$). As will be seen, the solutions found for the stress, strain
or displacement fields actually verify simpler, first or second-order PDEs
which can be used instead of (\ref{eq:Dop}).

\subsection{Averages, standard deviations and field distributions}
For any stress or strain component $a(\mathbf{x})$ produced by a loading
$\overline{a}$, normalized phase averages and standard deviations in phase
$\beta$ are defined as:
\begin{equation}
M^{(\beta)}(a)\equiv \langle a\rangle_{(\beta)}/\overline{a},\qquad
S^{(\beta)}(a)\equiv\sqrt{\langle a^2\rangle_{(\beta)}- \langle
a\rangle_{(\beta)}^2}/\overline{a}.
\end{equation}
The distribution (i.e.\ the histogram) of $a$ in the matrix $M$, $P_a(t)$, is
obtained by counting occurrences with the Dirac distribution $\delta$. All
individual averages in the matrix result from the definition $\langle
g(a)\rangle_{(1)}=\int {\rm d}t\,g(t)P_{a}(t)$, where:
\begin{equation}
\label{eq:distridef} P_{a}(t) \equiv
(1-f)^{-1}\int_{M}\!\,\text{d}^2\!x\, \delta(a(\mathbf{x})-t).
\end{equation}

\section{Material with anisotropy ratio $\alpha = 0$}
\label{sec:ellzero}
\subsection{Loading in pure shear}
\label{sec:k0PS} In this section, $\alpha=0$ with $\lambda=0$. A PS stress
loading $\overline{\sigma}=\langle\sigma_{\rm PS}\rangle$ is applied.

\subsubsection{Stress fields}
Using (\ref{eq:lambdamu}), $\lambda=0$ implies
$\sigma_\text{SS}=\sigma_{xy}\equiv 0$. Then, from stress equilibrium,
\begin{equation}
 \label{eq:formk0ps}
 \sigma_{xx}(x,y)=g(y), \quad \sigma_{yy}(x,y)=-g(x),
\end{equation}
where $g$ is an unknown 1-periodic function (using symmetry
\ref{eq:sym3}). Hence:
\begin{equation}
\label{eq:sigg} \sigma_\text{PS,m}=[g(y)\pm g(x)]/2,
\end{equation}
where the \textit{plus} (resp.\ \textit{minus}) sign applies to
$\sigma_\text{PS}$ (resp.\ $\sigma_\text{m}$). \emph{This sign convention for
PS loading, repeatedly employed hereafter, only holds for Sec.\ \ref{sec:k0PS}.
The opposite convention will apply in Secs.\ \ref{sec:k0SS} and
\ref{subsec:eqbxlam0}}.

\begin{figure}
\begin{center}
\epsfig{file=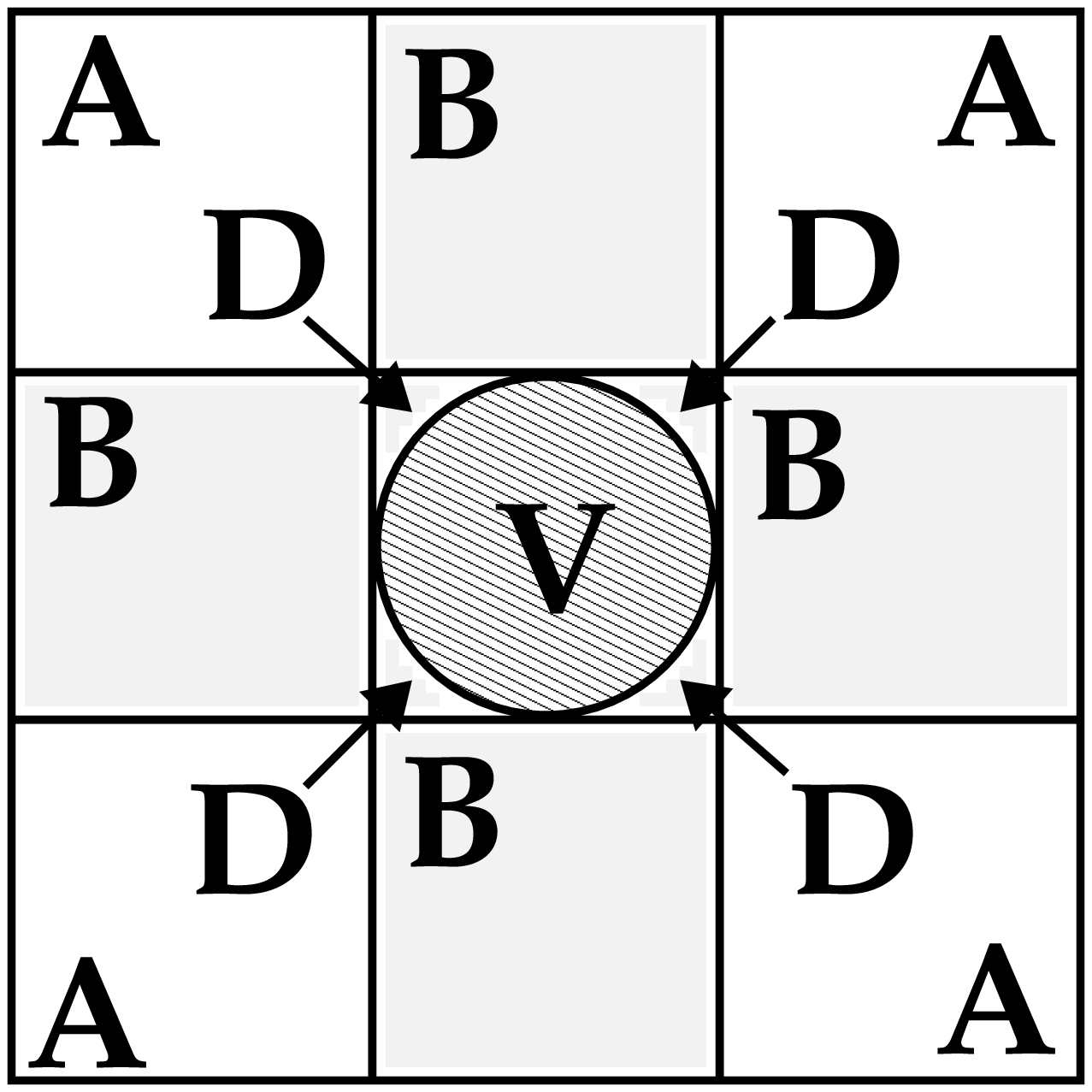,width=3.5cm}\,
(a)\hspace{0.5cm}\epsfig{file=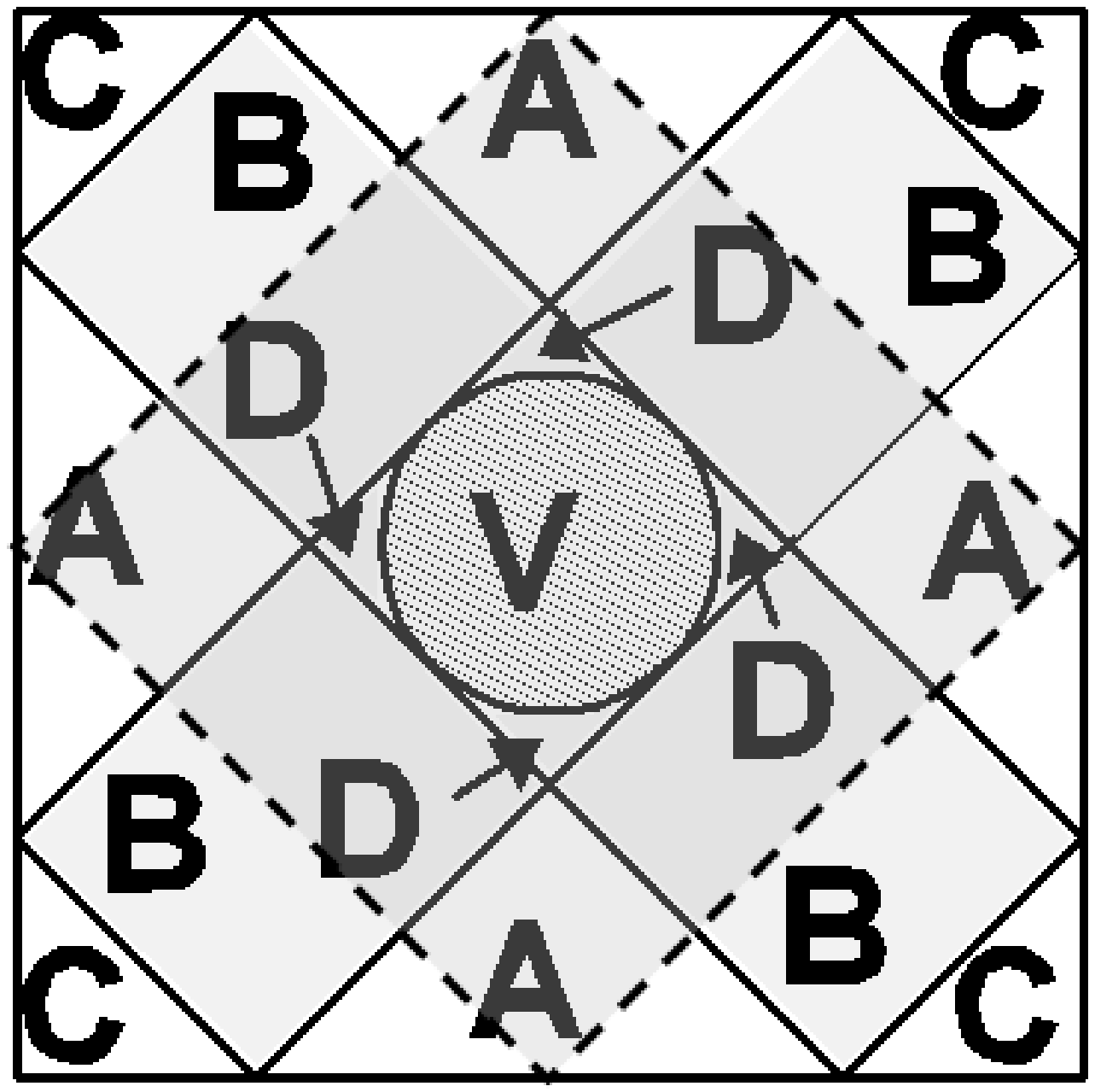,width=3.5cm}\,(b)
\begin{minipage}{12cm}
\caption{\label{fig:bands} Structure of field patterns in the unit
cell, in situations of infinite anisotropy. Figure (a): pattern for
$\alpha=0$. Figure (b): $\alpha=\infty$ (cf.\ Sec.\
\ref{sec:ellinfty}).}
\end{minipage}
\end{center}
\end{figure}

We note that each stress component $\sigma_{ii}$ (no summation) obeys a simple
first-order PDE, where $x=cst$ and $y=cst$ are characteristic lines, in
agreement with (\ref{eq:charact0}). The stress vanishes in the void so that
$g(x)\equiv 0$ for $x\in[-a,a]$. The structure of the solution is most easily
grasped referring to Fig.\ \ref{fig:bands}(a). It is organized in three types
of zones: (A), (B), and (D), separated by ``frontiers'' marked by solid lines.
The PS stress reduces to $\sigma_\text{PS}=g(x)/2$ or $g(y)/2$ in zone (B), to
$\sigma_\text{PS}=0$ in zone (D), and to $\sigma_\text{PS}=[g(x)+g(y)]/2$ in
zone (A). It thus vanishes in the square of length $2a$ consisting in the union
of (D) and of the void (V). This is more easily understood in terms of
characteristics. Since the transverse stress component $\sigma_{xy}$ is zero,
the void boundary conditions for the stress reduces to
$\sigma_{xx}=\sigma_{yy}=0$ at any point of the void-matrix interface. Hence,
each of these components is zero along one of the two families of
characteristics (vertical or horizontal) passing through a void. At the
intersection of characteristic lines passing through a void (i.e. region D),
all stress components must be zero. Thus, as far as stress is concerned, the
voids behave as \emph{square} voids. In particular, the effective modulus
(\ref{eq:PSmueff}) must be a natural function of $a$ rather than of $f$ (i.e.\
must not contain $\pi$ when expressed in terms of $a$).

The unknown function $g$ is obtained by minimizing the complementary elastic
energy functional (\ref{EffConst_u}), which is a strictly convex problem in the
subspace of periodic stress fields with vanishing $SS$ component (i.e.,
$\sigma_\text{SS}= 0$):
\begin{eqnarray}
 \label{energy}
&&(1-f)\langle u^{(1)}(\mbox{\boldmath $\sigma$})\rangle_{(1)} =
 \int_{-\frac{1}{2}\leq x, y\leq \frac{1}{2}}\left[
  \frac{1}{2\mu}\sigma_\text{PS}^2(x, y)
   +\frac{1}{2\kappa}\sigma_\text{m}^2(x, y)\right]\,{\rm d}x\, {\rm d}y,
   \nonumber\\
&&\qquad\qquad\qquad=\frac{1}{2}\left(\frac{1}{\mu}+\frac{1}{\kappa}\right)
           \int_a^{1/2}
g^2(x)\,{\rm d}x \nonumber +\left(\frac{1}{\mu}-\frac{1}{\kappa}\right)
   \left[\int_a^{1/2}
g(x)\,{\rm d}x\right]^2.
\end{eqnarray}
Functionally ex\-tre\-mi\-zing this integral with respect to $g$ under the
constraint
\begin{equation}
\label{eq:constrpureshear} \langle
\sigma_\text{PS}\rangle=\overline{\sigma}=2\int_a^{1/2}
\!\!\!\!\!\!g(x)\,{\rm d}x
\end{equation}
provides $g(x)\equiv \overline{\sigma}/(1-2a)$ for $x\in [-1/2,-a]\cup[a,1/2]$.
Denoting by $\theta_{[x_1,x_2]}(z)$ the characteristic function of the interval
$[x_1,x_2]$, we introduce $\chi(z)\equiv 1-\theta_{[-a,a]}(z)$, the
characteristic function of the domain $[-1/2,-a]\cup[a,1/2]$. Then,
$\sigma_\text{PS}$ and $\sigma_\text{m}$ are completely determined by:
\begin{equation}
\label{eq:sigsol1}\sigma_\text{PS,\,m}(x,y)=
\overline{\sigma}\frac{\chi(y)\pm\chi(x)}{2(1-2a)}.
\end{equation}
Referring to Fig.\ \ref{fig:bands}(a), $\sigma_{PS}$ meets its highest
(constant) value in zones (A), is of half this value in zones (B), and zero
elsewhere. The discontinuous stress pattern obtained here as a solution for
infinite anisotropy is of the type used by Drucker (1966) in his limit analysis
of the ideally plastic porous medium.

\subsubsection{Strain field}
\label{sub:PSk0strain} From (\ref{eq:lm}) and (\ref{eq:sigsol1}), the strain
components $\varepsilon_\text{PS}$ and $\varepsilon_\text{m}$ in the matrix
read:
\begin{equation}
\label{eq:k0PSepspamsm}\varepsilon_\text{PS}(x,y)=\overline{\sigma}
\frac{\chi(y)+\chi(x)}{4\mu(1-2a)},\qquad
\varepsilon_\text{m}(x,y)=\overline{\sigma}
\,m\,
\frac{\chi(y)-\chi(x)}{4\mu(1-2a)}.
\end{equation}
The relationship between $\overline{\sigma}$ and $\overline{\varepsilon}
=\langle\varepsilon_{\rm PS}\rangle$ is now obtained. To compute
$\overline{\varepsilon}$, we use $\mathbf{u}^*$. Expressions
(\ref{eq:k0PSepspamsm}) equivalently read [cf.\ equation (\ref{eq:a1e})]:
\begin{equation}
\label{eq:k0PSepsxxyy} \varepsilon_{xx}(x,y)=-\varepsilon_{yy}(y,x)=
\frac{1}{4\mu}\frac{\overline{\sigma}}{1-2a}[\left(1-m\right)\chi(x)
+\left(1+m\right)\chi(y)].
\end{equation}
In PS loading, the admissibility (i.e., compatibility) conditions in
(\ref{EffW}) imply that $\partial_x
u^*_x=\varepsilon_{xx}-\overline{\varepsilon}$ and $\partial_y
u^*_y=\varepsilon_{yy}+\overline{\varepsilon}$. The displacement component
$u^*_x$ (resp. $u^*_y$) is odd and 1-periodic wrt.\ $x$ (resp. $y$, see Eqs.\
\ref{eq:sym3}). This requires $\mathbf{u}^*$ to be tangent to the boundary of
the unit cell: $u^*_x(\pm 1/2,y)\equiv u^*_y(x,\pm 1/2)\equiv0$. Hence, the
previous Eqs.\ are integrated as:
\begin{equation}
\label{eq:uxuy} u_x^*(x,y)=\int^x_{-1/2} {\rm d}x'\,
\left[\varepsilon_{xx}(x',y)-\overline{\varepsilon}\right],\quad
u_y^*(x,y)=\int^y_{-1/2} {\rm d}y'\,
\left[\varepsilon_{yy}(x,y')+\overline{\varepsilon}\right].
\end{equation}
Furthermore, the periodicity condition $u^*_x(1/2,\cdot)\equiv 0$ yields:
\begin{equation}
\int^{1/2}_{-1/2} {\rm d}x\,
\left[\varepsilon_{xx}(x,y)-\overline{\varepsilon}\right]=0.
\label{eq:LineIntegration}
\end{equation}
Choosing, e.g., $y=1/2$ (in the matrix) and inserting
(\ref{eq:k0PSepsxxyy}) in (\ref{eq:LineIntegration}) provides the
desired relation between $\overline{\sigma}$ and
$\overline{\varepsilon}$:
\begin{equation}
\label{eq:eps0sig0}
\overline{\sigma}=\frac{2\mu(1-2a)}{1+(m-1)a}\overline{\varepsilon}
\end{equation}
Since $m>0$ and $0\leq a\leq 1/2$, the denominator of (\ref{eq:eps0sig0}) is
always $>0$.

The still unknown $\varepsilon_{xy}\equiv\varepsilon_\text{SS}$ is computed
from the admissibility (i.e., compatibility) conditions in (\ref{EffW}), and
from the expressions of $u^*_x$ and $u^*_y$ in (\ref{eq:uxuy}). We obtain for
$(x,y)$ in the matrix:
\begin{equation}
\label{eq:k0PSepe}
 \varepsilon_{xy}(x,y)= \frac{1}{2}\left[\int_{-1/2}^x {\rm d}x'
  \frac{\partial}{\partial y}\varepsilon_{xx}(x', y)
+ \int_{-1/2}^y {\rm d}y'
  \frac{\partial}{\partial x}\varepsilon_{yy}(x, y')\right].
\end{equation}
Inserting the solutions (\ref{eq:k0PSepsxxyy}), the integration is first
carried out in the LLQ for integration paths in the matrix. The result is
extended to the whole unit cell appealing to identities (\ref{eq:a1f},g). For
instance, in the URQ:
\begin{equation}
 \label{eq:k0PSepspe}
\varepsilon_\text{SS}(x,y)=\frac{\overline{\varepsilon}}{4}\frac{1+m}{1+(m-1)a}
\left[\left(x-\frac{1}{2}\right)\delta(y-a) -
\left(y-\frac{1}{2}\right)\delta(x-a)\right],
\end{equation}
where the Dirac distributions stem from the discontinuities in the
displacements (\ref{eq:uxuy}) due to the $\chi$ functions in
(\ref{eq:k0PSepsxxyy}). Hence, in PS loading, $\varepsilon_\text{SS}$ is
localized on the four ``frontier'' lines in the unit cell. Points $(x,y)=(\pm
a,0)$ and $(0,\pm a)$ are singular \emph{hot spots}. At each tangency point of
a ``frontier'' line with the void boundary, there occurs a jump of
$\varepsilon_\text{SS}$ along the line. E.g., in the vicinity of $(a,0)$,
(\ref{eq:k0PSepspe}) and the symmetry
$\varepsilon_{xy}(x,-y)=-\varepsilon_{xy}(x,y)$ provide:
\begin{equation}
\label{eq:exysingexample} \varepsilon_\text{SS}(x,y)\simeq
\frac{\overline{\varepsilon}}{8}\frac{1+m}{1+(m-1)a}\delta(x-a)\,\mathop{\text{sign}}y.
\end{equation}

\subsubsection{Displacement field and ``hot spots''.}
From (\ref{eq:uxuy}) and identity $u^*_x(x,y)=-u^*_x(-x,-y)$, the displacement
$\mathbf{u}^*$ corresponding to the above strains reads in the URQ ($\theta$ is
the Heaviside function):
\begin{eqnarray}
\label{eq:k0PSuxFull}
u_x(x,y)&=&\frac{\overline{\varepsilon}}{1+(m-1)a}
\left\{\left(\frac{1}{2}-x\right)
\left[(m-1)a+\frac{1}{2}(m+1)\theta(a-y)\right]\right.\nonumber\\
&&{} \left. -\frac{1}{2}(m-1)(a-x)\theta(a-x)\right\}.
\end{eqnarray}
Moreover, $u^*_y(x,y)=-u^*_x(y,x)$. Again referring to Fig.\
\ref{fig:bands}(a), $\mathbf{u}^*$ is a linear function of $x$, $y$ in (A), (B)
and (D). Along the ``frontiers'' between (A) and (B) or between (B) and (D),
its component normal to the ``frontier'' is continuous, whereas its tangential
component is discontinuous. For instance, on the ``frontier'' $y=a$ for
$a<x<1/2$, $u_x^*$ undergoes a jump $[[u_x^*]]_y\equiv
u_x^*(x,a^+)-u_x^*(x,a^-)$ given by:
\begin{equation}
\label{jump} [[u_x^*]]_y(x,a)=-u_1 (1/2-x), \qquad
u_1\equiv\frac{\overline{\varepsilon}(1+m)}{2[1+(m-1)a]}.
 \end{equation}
\begin{figure}
\begin{center}
\epsfig{file=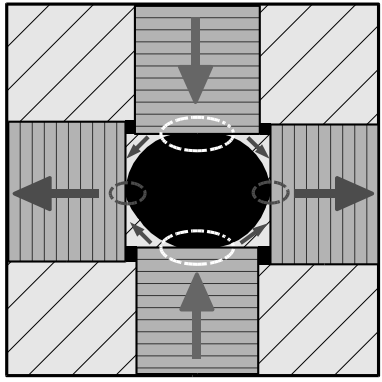,width=5cm}\\
\begin{minipage}{12cm}
\caption{\label{fig:taquin} Structure of the deformed matrix and of
$\mathbf{u^*}$ (arrows) for an anisotropy ratio $\alpha=0$, in the
particular case of equal bulk and shear moduli $\kappa=\mu$. Loading
is PS. Voids are in black. White (resp.\ grey) elliptic rings tag
zones of extreme matter separation (resp.\ crushing).}
\end{minipage}
\end{center}
\end{figure}
As expected, the displacement field can develop discontinuities along
characteristic lines. This feature is reminiscent of rigid, ideal\-ly\--plastic
bodies in isotropic two-dimensional materials where discontinuities may only
occur tangentially to slip lines (Kachanov, 1974, prop.\ 39.4). The structure
of the full deformed configuration is schematized in Fig.\ \ref{fig:taquin}:
overall deformation occurs by block sliding. Fig.\ \ref{fig:taquin} is drawn in
the particular case of equal bulk and shear moduli, $\kappa=\mu$ ($m=1$). Then,
$\mathbf{u^*}$ vanishes strictly in zone (A) --- but is finite there if
$\kappa\not=\mu$ --- and is aligned with the axes in (B).

The expression of $\mathbf{u}$ in (D) clarifies the nature of the hot spots:
\begin{equation}
\label{eq:uzoneD} \mathbf{u}(x,y)= u_1(\mathop{\text{sign}}x,
-\mathop{\text{sign}}y), \qquad -a<x,y<a,
\end{equation}
We stress that this incomplete expression, valid for $-a<x,y<a$ only, does not
allow one to recover the singularity (\ref{eq:exysingexample}) of
$\varepsilon_{xy}$ by taking a derivative. Since $\mathbf{u}$ is constant, save
for orientation changes, in the four quadrants of zone (D), the hot spots are
either points of extreme matter separation (at $(x,y)=(0,\pm a)$ for
$\overline{\varepsilon}>0$; white elliptic markings in Fig.\ \ref{fig:taquin})
or of matter crushing (at $(x,y)=(0,\pm a)$; dark markings).

Fig.\ \ref{fig:taquin} also makes conspicuous four voided squares (in black)
generated by the block-sliding pattern, at the intersections $(x,y)=(\pm a,\pm
a)$ of the ``frontier'' lines (the phenomenon is most remarkable for $m=1$).
The immediate vicinity of \emph{each} of these points is divided into four
regions where the displacement vector locally takes on distinct values. For
instance, around $(x,y)=(a,a)$, we have $\mathbf{u^*}(a^+,a^+)=( u_0,-u_0)$,
$\mathbf{u^*}(a^+,a^-)=( u_0+\Delta u,-u_0)$, $\mathbf{u^*}(a^-,a^+)=(u_0,-u_0-
\Delta u)$, $\mathbf{u^*}(a^-,a^-)=(u_0+\Delta u,-u_0-\Delta u)$, where
\begin{equation}
u_0=2u_1[(1-m)/(1+m)] a(a-1/2),\qquad \Delta u=u_1(1/2-a).
\end{equation}
The quantity $\Delta u$ represents the size of the voided square, to first
order in $\overline{\varepsilon}$. Remark that in terms of $\overline{\sigma}$,
it reads $\Delta u=\overline{\sigma}(1+m)/(8\mu)$, a void-independent
expression (the dimensioning factor is the cell size $L=1$).

\subsubsection{Distributions, moments and effective shear modulus $\widetilde{\mu}(f)$}
\label{sec:alpha0pseff} 
\begin{figure}
\begin{center}
\epsfig{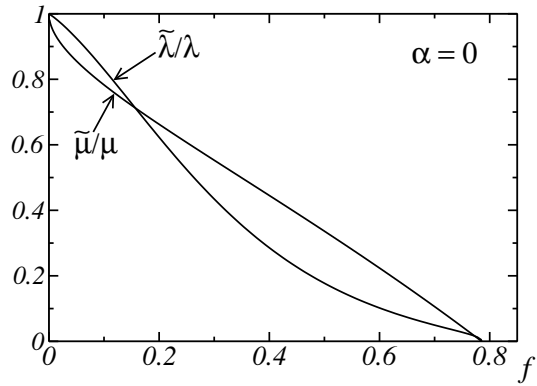}
\begin{minipage}{12cm}
\caption{\label{fig:effectiveModulivsfLzero} Normalized shear
effective moduli $\widetilde{\mu}/\mu$ and
 $\widetilde{\lambda}/\lambda$ vs.\ void concentration
$f$, for an incompressible matrix with anisotropy ratio
$\alpha=\lambda/\mu=0$. Both moduli vanish at the close packing
threshold $f=f_c=\pi/4\simeq 0.78$. }
\end{minipage}
\end{center}
\end{figure}

In the PS case, the distributions
$P_{\varepsilon_z}(t)$ or $P_{\sigma_z}(t)$ consist uniquely of a sum of
Dirac-type components, plus unusual singular components at
$\varepsilon_z\to\pm\infty$, due to the Dirac singularities in the fields. The
latter cannot be accounted for by probability densities unless inconvenient
limiting processes are employed. This may constitute a limitation of the use of
distributions in a localizing regime. However, the first and second moments are
readily computed. The effective shear modulus $\widetilde{\mu}$ is read from
(\ref{eq:eps0sig0}). With $a\equiv\sqrt{f/\pi}<1/2$, we have:
\begin{subequations}
\label{eq:momentsEll0PS}
\begin{equation}
\label{eq:PSmueff}
\frac{\widetilde{\mu}}{\mu}=\frac{1-2a}{1+(m-1)a}=1-(1+m)(f/\pi)^{1/2}+O(f),
\end{equation}
\begin{equation}
\label{eq:PSfieldmeans}
M^{(1)}(\varepsilon_\text{PS})=\frac{1-2a}{(1-f)[1+(m-1)a]},\quad
  M^{(2)}(\varepsilon_\text{PS})=\frac{(m+1)a}{f[1+(m-1)a]},
\end{equation}
\begin{equation}
S^{(1)}(\varepsilon_\text{PS})= \frac{\sqrt{(1-2a)[(1+f)a-f]}}
     {(1-f)[1+(m-1)a]}, \quad S^{(1)}(\sigma_\text{PS})=
     \frac{\sqrt{(1+f)a-f}}{(1-f)\sqrt{1-2a}},
\end{equation}
\begin{equation}
S^{(1)}(\varepsilon_\text{m})=
\frac{m\sqrt{a(1-2a)}}{[1+(m-1)a]\sqrt{1-f}}, \quad
S^{(1)}(\sigma_\text{m})=\sqrt{\frac{a}{(1-f)(1-2a)}},
\end{equation}
\begin{equation}
\label{eq:S1ss} S^{(1)}(\varepsilon_\text{SS})=\infty, \quad
S^{(1)}(\sigma_\text{SS})=0.
\end{equation}
\end{subequations}
The mean strain in the pore stems from
$M^{(2)}(\varepsilon_\text{PS})=[1-(1-f)M^{(1)}(\varepsilon_\text{PS})]/f$.
Moreover,
$S^{(1)}(\sigma_\text{PS})=(\mu/\widetilde{\mu})S^{(1)}(\varepsilon_\text{PS})$.
The Dirac singularities are responsible for the transverse variance
$S^{(1)}(\varepsilon_\text{SS})$ being infinite. Also, remark that
$M^{(2)}(\varepsilon_\text{PS})$ blows up as $f^{-1/2}$ when $f\to 0$. The
curve $\widetilde{\mu}(f)/\mu$ is displayed in Fig.\
\ref{fig:effectiveModulivsfLzero} in the incompressible case $m=0$. The power
singularity at $f=0$ with infinite negative slope, due to the $f^{1/2}$ term in
the dilute expansion of (\ref{eq:PSmueff}), indicates that an infinitesimal
void dramatically weakens the medium. This is reminiscent of the situation
encountered in plasticity (see Introduction). In the linear material considered
here, it is a direct consequence of hyperbolicity. Each of the two components
$\sigma_{xx}$ and $\sigma_{yy}$ have constant values along one of the two
families of characteristics. Due to the boundary conditions at the void-matrix
interface, the parallel component $\sigma_\textnormal{PS}$ is smaller by half
in region (B) than it is in (A). Hence, each void lowers the stress field over
large regions in the material, which are projections of the voids along the
characteristic directions and involve \emph{infinitely long distances}. The
normalized standard deviations $S^{(1)}(\varepsilon_\text{PS})$,
$S^{(1)}(\varepsilon_\text{m})$, $S^{(1)}(\sigma_\text{PS})$ and
$S^{(1)}(\sigma_\text{m})$ all behave as $(f/\pi)^{1/4}$ as $f\to 0$, which
indicates that they grow more rapidly with $f$ than in a linear isotropic
medium. The effective modulus $\widetilde{\mu}(f)$ vanishes \emph{linearly}
with $(f_c-f)$ for $f$ near $f_c=\pi/4$, the void close packing fraction (see
the comment in Sec.\ \ref{sec:alpha0ssefmod}), along with the moments of the
strain components in the matrix in the loading direction
($M^{(1)}(\varepsilon_\text{PS})$ $=$ $S^{(1)}(\varepsilon_\text{PS})$ $=0$).

\subsection{Loading in simple shear (incompressible case only)}
\label{sec:k0SS} According to Sec.\ \ref{sec:ohdc}, SS loading goes
along with $\alpha=0$ and $\mu=\infty$. Matrix incompressibility renders the problem hyperbolic and is
assumed for simplicity ($\kappa=\infty$,
$m=0$).

\subsubsection{Displacement, strain and stress fields}
 \label{sec:k0SSdispl}
Taking $\mu=\infty$ under finite stress implies
$\varepsilon_{xx}-\varepsilon_{yy}=0$. Incompressibility then requires
$\varepsilon_{xx}=\varepsilon_{yy}=0$. The only non-zero component is
$\varepsilon_{xy}$, and a variational calculation is simpler to carry out wrt.\
the strain rather than to the stress. Condition
$\varepsilon_{xx}=\varepsilon_{yy}=0$ implies $u^*_x(x,y)=u^*_x(y)$ and
$u^*_y(x,y)=u^*_y(x)$. By symmetry, see \ref{eq:a2c}, $u^*_x(z)=u^*_y(z)\equiv
G(z)-\overline{\varepsilon}\,z$, where $G$ is unknown. Introducing the
derivative $g\equiv G'$, one obtains:
\begin{equation}
\label{eq:epsxy}
\varepsilon_{xy}(x, y)=[g(x)+g(y)]/2.
\end{equation}

\begin{figure}
\begin{center}
\rotatebox{0}{\includegraphics[width=
12cm]{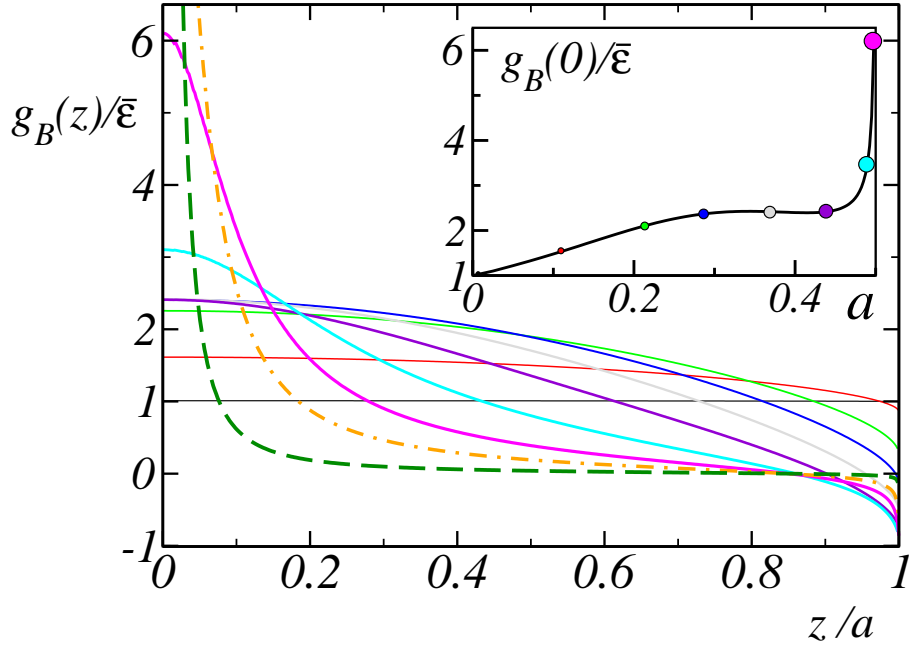}}
\begin{minipage}{12cm}
\caption{\label{fig:functiongb} Plots of the solution $g_B(z)$ of
(\ref{eq:exactGlobalMin1}) as a function of $z/a$, $0<z/a<1$, for values of the
void radius (from bottom to top at $z=0$) $a=0.025$ (black), $0.124$ (red),
$0.249$ (green), $0.275$ (blue), $0.373$ (gray), $0.435$ (violet), $0.485$
(cyan), $0.498$ (rose), $0.4995$ (orange), $0.499985$ (dark green). Color
online. Line and symbol thicknesses increase with $a$.}
\end{minipage}
\end{center}
\end{figure}

The energy of the unit cell (\ref{EffW}) is expressed as an integral over the
matrix phase $M$ (represented by the unit square minus the void):
\begin{equation}
(1-f)\langle w^{(1)}\rangle_{(1)}
 =2\lambda\int_M{\rm d}^2\!x\,\mathsf{\varepsilon}_{xy}^2
 =\frac{\lambda}{2}\Biggl\{
\int_{\vphantom{\biggl[}[-\frac{1}{2},\frac{1}{2}]
\times[-\frac{1}{2},\frac{1}{2}]}\hspace{-3.5em}{\rm d}x\,{\rm d}y
-\int_V{\rm d}x\,{\rm d}y \biggr\}[g(x)+g(y)]^2.
\end{equation}
After using identity (\ref{eq:a2f}), we expand it into separate integrals over
the intervals $[0,a]$ and $[a,1/2]$, and we split $g$ into independent
functions $g_A$ and $g_B$ supported by these intervals. Subscripts $A$, $B$
refer to the zones in Fig.\ \ref{fig:bands}(a):
\begin{equation}
\label{eq:geb0}
g(z)=g_B(z)\theta_{[0,a]}(z)+g_A(z)\theta_{[a,1/2]}(z).
\end{equation}
The strain energy (\ref{EffW}) is functionally minimized wrt.\ $g_A$, $g_B$ under the
constraint:
\begin{equation}
\label{constraink0SS}
\langle\varepsilon_{xy}\rangle\equiv\overline{\varepsilon}=2\left[\int_a^{1/2}g_A(z)\,{\rm d}z
+\int_0^ag_B(z)\,{\rm d}z\right].
\end{equation}
The system obtained determines $g_A$ as a constant. Introducing
$\rho(z)\!\equiv\!\!\sqrt{a^2-z^2}$, it provides an integral equation for
$g_B(z)$:
\begin{subequations}
\label{eq:systgb}
\begin{eqnarray}
\label{eq:epsaconst}
g_A(z)&\equiv& g_B(a),\qquad z\in[a,1/2],\\
\label{eq:exactGlobalMin1}
2\int_{\rho(z)}^a\hspace{-0.3cm}g_B(y){\,{\rm d}y} &=& \overline{\varepsilon}
+2 a\, g_B(a) +\left[2\rho(z)-1\right]
  g_B(z),\qquad z\in[0,a].
\end{eqnarray}
\end{subequations}
In particular, for $z=a$, $g_B(a)$ is expressed in terms of $g_B(z)$,
$z\in[0,a[\,$:
\begin{equation}
\label{eq:sumEpsB}
2\int_0^a {\rm d}z\,g_B(z)=\overline{\varepsilon}+\left(2a-1\right)g_B(a).
\end{equation}
Taking $z=0$ in (\ref{eq:exactGlobalMin1}) yields a relationship between
$g_B(0)$ and $g_B(a)$.
\begin{equation}
\label{eq:gB0gBa} g_B(0)=[\overline{\varepsilon}+2a\,g_B(a)]/(1-2a).
\end{equation}
Eqn.\ (\ref{eq:exactGlobalMin1}) can be transformed into a
second-order differential equation
with no simple analytical
solution.\footnote{With $Q(x)\equiv\int_0^x g_B(y)\,{\rm d}y$,
(\ref{eq:exactGlobalMin1}) relates linearly $Q(a-\rho(z))$ to $Q'(z)$.
Differentiating it wrt.\ $z$ yields another equation connecting
 $Q'(a-\rho(z))$, $Q'(z)$ and $Q''(z)$.
The substitution $y=\rho(z)$ admits $z=\rho(y)$ as an inverse, and turns the
first equation into one linking $Q(z)$ to $Q'(a-\rho(z))$ and the second one
into a differential equation for $Q(z)$.} Instead we express the strain, the
displacement and the stress in terms of $g_B$, and we directly deduce from
(\ref{eq:exactGlobalMin1}) some asymptotic results for the fields and for their
distributions. We also compute numerically the whole solutions, the
distributions, and the effective moduli by discretizing
(\ref{eq:exactGlobalMin1}) into a linear system. The solution for $g_B(z)$ is
represented v.s.\ $0<z/a<1$ for $a\in[0,1/2]$ in Fig.\ \ref{fig:functiongb}.
The inset displays $g_B(0)$ as a function of $a$. At $a$ fixed, the highest
values lie at $z=0$, the strain being higher on the cartesian axes of the unit
cell. Two regimes are encountered as $a$ increases: the strain first develops
in zone (B) and concentrates around the axes; then, for $a\gtrsim 0.4$ (i.e.\
$f\gtrsim 0.5$), $g_B(0)$ blows up as $(a_c-a)^{-1/2}$ near $a_c=1/2$ (see
Sec.\ \ref{sec:scaling}), while the strain localizes on the cartesian axes. In
the limit, $g_B(z)\propto \delta(z)$. The field enhancement between voids near
close packing is boosted by anisotropy. The close packing void fraction
$f_c=\pi/4$ is, for periodic media, tantamount to the void mechanical
percolation threshold in random porous media (e.g., Torquato 2002). When $a\to
a_c$, $g_B(z)$ becomes discontinuous, with $g_B(a)=-1$ and $g_B(a^{-})=0$, see
Fig.\ \ref{fig:functiongb}.

The solution in terms of $g_B$ is as follows. With (\ref{eq:epsaconst}), $g(z)$
in (\ref{eq:geb0}) becomes completely determined in terms of $g_B(z)$,
$z\in[0,a]$. In the URQ:
\begin{equation}
\label{eq:epsxyepsb}
\varepsilon_{xy}(x,y)=\frac{g_B(a)}{2}\left[\theta_{[a,1/2]}(x)+\theta_{[a,1/2]}(y)\right]
+\frac{1}{2}[g_B(x)\theta_{[0,a]}(x)+g_B(y)\theta_{[0,a]}(y)].
\end{equation}
The integral equation (\ref{eq:exactGlobalMin1}) admits a continuous solution.
Then, $g_B$ and in turn $\varepsilon_{xy}$, $\mathbf{u}$ and $\mathbf{u^*}$ are
continuous, too. The continuity and oddness wrt.\ $x$ (resp. $y$) of $u^*_x$
(resp. $u^*_y$) implies
 $u_x^*(0,y)\equiv u_y^*(x,0) \equiv 0$. Hence, in the URQ:
\begin{equation}
\label{eq:ustar} u_x^*(x,y)=\int_0^y{\rm
d}z\,g(z)-\overline{\varepsilon}y,\qquad
u_y^*(x,y)=u_x^*(y,x),\qquad (0<x, y<1/2).
\end{equation}
These expressions are extended to the unit cell using
(\ref{eq:a2a}-\ref{eq:a2c}). From (\ref{eq:a2d}) the transverse components
$\sigma_{xx}$ and $\sigma_{yy}$ are odd in $x$ and $y$, so that periodicity
requires $\sigma_{xx}(1/2,y)\equiv\sigma_{yy}(x,1/2) \equiv 0$. With
$\sigma_{xy}(x,y)=2\lambda\,\varepsilon_{xy}(x,y)$ given by
(\ref{eq:epsxyepsb}), we deduce from stress equilibrium in the URQ of the
matrix:
\begin{equation}
\label{eq:sigmak0SS}
\sigma_{xx}(x,y)=\sigma_{yy}(y,x)=
\int_x^{1/2} {\rm d}\!\,x'\, \partial_y \sigma_{xy}(x',y)
=(\lambda/2)(1-2x)g'(y).
\end{equation}
Then in this region:
$$\sigma_\text{m,PS}(x,y)=(\lambda/4)[(1-2x)g'(y)\pm(1-2y)g'(x)],$$
where the \textit{plus} (resp.\ \textit{minus}) sign applies to
$\sigma_\text{m}$ (resp.\ $\sigma_\text{PS}$). Hence, because
$g_B(z)=\text{const.}$ for $z\in[a,1/2]$, the stresses $\sigma_\text{m}$ and
$\sigma_\text{PS}$ vanish in zones (A) of Fig.\ \ref{fig:bands}(a). In zones
$(B)$, $\sigma_\text{m}$ and $\sigma_\text{PS}$ are closely related, with:
\begin{subequations}
\label{eq:sigmmpsB}
\begin{eqnarray}
\sigma_\text{m}(x,y)&=&\sigma_\text{PS}(x,y)
=(\lambda/4)(1-2x)g_B'(y)\quad\text{ for }
\left\{
\genfrac{}{}{0pt}{1}{a<x<1/2}{0<y<a}\right.,\\
\label{eq:sigmmpsBb} \sigma_\text{m}(x,y)&=&-\sigma_\text{PS}(x,y)
=(\lambda/4)(1-2y)g_B'(x)\quad\text{ for } \left\{
\genfrac{}{}{0pt}{1}{0<x<a}{a<y<1/2}\right..
\end{eqnarray}
\end{subequations}
Taking the derivative of (\ref{eq:exactGlobalMin1}) with respect to $z$, we
deduce
\begin{equation}
\label{eq:gBder}
g_B'(z)=-2z\frac{g_B(z)+g_B\left(\rho(z)\right)}{\rho(z)[1-2\rho(z)]},
\end{equation}
so that $g_B'(0)=0$. Hence besides zones (A), regions of weak
$\sigma_\text{m}$ and $\sigma_\text{PS}$ lie along the axes $x=0$ or
$y=0$, and along the axes $x=\pm 1/2$, $y=\pm 1/2$.

Near the ``frontiers'' in Fig.\ \ref{fig:bands}(a), $\sigma_\text{m}$ and
$\sigma_\text{PS}$ blow up. Indeed, let us focus on the ``frontier'' $y=a$.
From (\ref{eq:gBder}), we deduce for $y\lesssim a$
\begin{equation}
\label{eq:singya}
g_B'(y)\simeq-\sqrt{2a}[g_B(a)+g_B(0)](a-y)^{-1/2},
\end{equation}
and the latter stresses behave as $d^{-1/2}$, where $d$ is the distance to the
``frontier'' line. The variances involving an integral with integrand $\propto
\sigma^2$, this implies:
\begin{equation}
\label{eq:coneinfty}
S^{(1)}(\sigma_\text{PS})=S^{(1)}(\sigma_\text{m}) = \infty.
\end{equation}
These divergences are the only ones encountered in the means and variances. The
mean stress readily derives from Eqs.\ (\ref{eq:systgb}-\ref{eq:sumEpsB}) and
(\ref{eq:epsxyepsb}):
\begin{equation}
\label{eq:meanstressxy}
\overline{\sigma}=\langle\sigma_{xy}\rangle=8\lambda\left[
\int_0^{1/2}\hspace{-0.45cm}{\rm d}x\int_0^{1/2}\hspace{-0.45cm}{\rm
d}y-\int_0^a\hspace{-0.25cm}{\rm
d}x\int_0^{\rho(x)}\hspace{-0.45cm}{\rm d}y \right]\varepsilon_{xy}
=\lambda\overline{\varepsilon}\left[1+g_B(a)/\overline{\varepsilon}\right],
\end{equation}
\begin{equation}
\label{eq:lameff1} \hspace{-0.1em}\text{The effective shear modulus
is therefore}\quad
\widetilde{\lambda}=(\lambda/2)\left[1+g_B(a)/\overline{\varepsilon}\right].
\end{equation}

\subsubsection{Moments and effective shear modulus $\widetilde{\lambda}(f)$ in the dilute limit}
\label{sec:alpha0ssefmod} Series expansions are now carried out for
$a\ll 1$. Using the Taylor expansion:
\begin{equation}
\label{eq:taylor} g_B(a z)=
\overline{\varepsilon}\sum_{n=0}^{\infty}q_n(z)a^n\qquad (0<z<1),
\end{equation}
we solve Eqn.\ (\ref{eq:exactGlobalMin1}). Note that $a$ being a parameter, we
should have denoted $g_B(z)$ by $g_B(z;a)$, so that in the above equation,
$g_B(a z)$ would stand for $g_B(a z;a)$, expanded in powers of $a$. The
following recursion is obtained:
\begin{equation}
\frac{q_{n+1}(z)}{2}=q_n(1)+\sqrt{1-z^2}q_n(z)
-\int_{\sqrt{1-z^2}}^1 \text{d}\!y\,q_n(y),\qquad q_0(z)\equiv 1.
\end{equation}
Numerically, we find that the series (\ref{eq:taylor}) converges for
$a<R_a\simeq 0.4$. In the convergence region, this solution matches the
numerical one obtained in the previous section. We carry out analytically the
recursion up to fourth order to estimate $g_B(az)/\overline{\varepsilon}$. The
integrations are readily performed with a symbolic calculator. Using the result
with $z=1$ provides:
\begin{equation}
\label{eq:expana}
g_B(a)/\overline{\varepsilon}=1-2\pi a^2-\frac{64}{3}a^3+2(\pi^2-6\pi-8)a^4+O(a^5).
\end{equation}
With expansion (\ref{eq:expana}), we find in the dilute limit:
\begin{subequations}
\label{eq:momentsEll0SS}
\begin{equation}
\label{eq:lam0} \frac{\widetilde{\lambda}}{\lambda}=
1-f-\frac{32f^{3/2}}{3\pi^{3/2}}
+\left(1-\frac{6}{\pi}-\frac{8}{\pi^2}\right)f^2+ O(f^{5/2}),
\end{equation}
\begin{equation}
M^{(1)}(\varepsilon_{\text{SS}})=1-\frac{32\,f^{3/2}}{3\pi^{3/2}}
+O(f^2),\quad M^{(2)}(\varepsilon_\text{SS})=
1+\frac{32\,f^{1/2}}{3\pi^{3/2}} +O(f),
\end{equation}
\begin{equation}
S^{(1)}(\varepsilon_\text{SS})=\frac{4\sqrt{2}\,f^{3/4}}{\sqrt{3}\pi^{3/4}}
+O(f^{5/4}), \quad
S^{(1)}(\sigma_\text{SS})=\frac{4\sqrt{2}\,f^{3/4}}{\sqrt{3}\pi^{3/4}}
+O(f^{5/4}),
\end{equation}
\begin{equation}
S^{(1)}(\varepsilon_\text{PS})=0,\quad
S^{(1)}(\sigma_\text{PS})=\infty,\quad
S^{(1)}(\varepsilon_\text{m})= 0,\quad S^{(1)}(\sigma_\text{m}) =
\infty.
\end{equation}
\end{subequations}
A plot of $\widetilde{\lambda}(f)/\lambda$ obtained from the full numerical
solution of $g_B(z)$ is displayed in Fig.\ \ref{fig:effectiveModulivsfLzero}.
Compared to $\widetilde{\mu}(f)/\mu$, the less singular character of the
solution goes along with a harder material at small porosities. But
$\widetilde{\lambda}$ falls down to zero much faster than $\widetilde{\mu}$ for
higher porosities, excepted near the close-packing threshold.

\subsubsection{Scaling in the close packing limit}
\label{sec:scaling} Day \textit{et al.} (1992) examined the linear elastic
behavior of an isotropic material containing a honeycomb lattice of voids: the
effective compressibility and shear moduli (the overall medium is isotropic)
vanish as $(f_c-f)^{1/2}$ near the void packing threshold. A similar behavior
is found here in shear for the anisotropic material.

\begin{figure}[htbp]
\begin{center}
\rotatebox{-90}{\includegraphics[width=7cm]{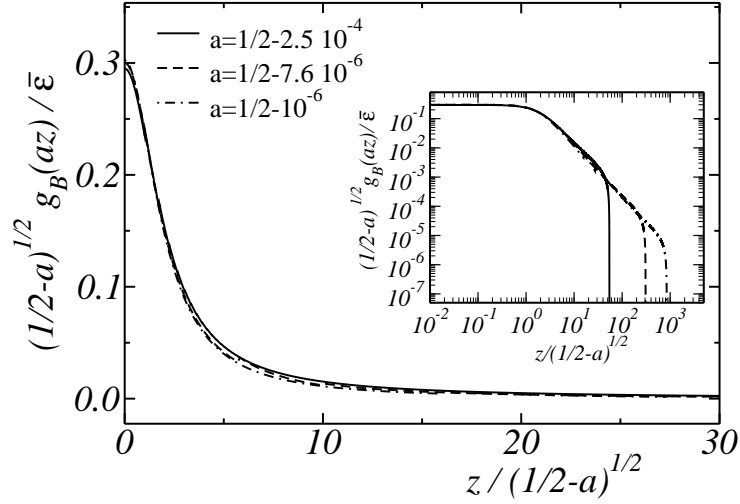}}
\begin{minipage}{12cm}
\caption{\label{fig:gscaling} Data collapse illustrating the scaling
properties of the function $g_B(z)$, for the 3 different values of
the radius $a$ indicated in the legend, near the close packing value
$a_c=1/2$. Inset: same plot in log.-log. scale; the curves break
down when $g_B(az)$ become negative.}
\end{minipage}
\end{center}
\end{figure}

With the numerical solution of (\ref{eq:exactGlobalMin1}) near $f_c$, we
observe that the function $g_B(z)$ obeys a scaling of the form
\begin{equation}
\label{eq:closePacking} g_B(a z)\sim \overline{\varepsilon}(f_c-f)^{-1/2}
 \widetilde{g}\left(z(f_c-f)^{-1/2}\right), \quad f\to f_c,
\end{equation}
where the master curve $\widetilde{g}(z)\sim$ const.\ for $z\ll 1$ and
$\widetilde{g}(z)\sim z^{-\tau}$ for $z\gg 1$, where $\tau$ is an exponent
close to 2. Fig.\ \ref{fig:gscaling} provides an illustration for three
different values of $a$ near $a_c=1/2$.

Two consequences are drawn from (\ref{eq:closePacking}): first, the scaling
shows that the strain is concentrated in a band of width $\xi\sim
a(f_c-f)^{1/2}$, vanishing as the close-packing threshold is approached. Next,
relation (\ref{eq:closePacking}) used at $z=0$ together with (\ref{eq:gB0gBa})
and (\ref{eq:lameff1}) provides near $f_c$:
\begin{subequations}
\begin{equation}
\widetilde{\lambda}/\lambda\sim(f_c-f)^{1/2},
\end{equation}
as for an isotropic material. The means are readily computed. Moreover, the
variances are obtained with the help of the surface integral in the matrix
$\int_M {\rm d}x\,{\rm
d}y\,\varepsilon_\text{SS}^2(x,y)=\overline{\varepsilon}^2
[1+g_B(a)/\overline{\varepsilon}]/2$. The following behaviors near the
close-packing threshold ensue:
\begin{equation*}
M^{(1)}(\varepsilon_\text{SS})\sim \left(f_c-f\right)^{1/2},\hfill
S^{(1)}(\varepsilon_\text{SS})\sim\left(f_c-f\right)^{1/4}, \hfill
S^{(1)}(\sigma_\text{SS})\sim\left(f_c-f\right)^{-1/4},
\end{equation*}
\begin{equation}
S^{(1)}(\varepsilon_\text{PS})=S^{(1)}(\varepsilon_\text{m}) \equiv
0, \qquad S^{(1)}(\sigma_\text{PS})=S^{(1)}(\sigma_\text{m}) \equiv
\infty.
\end{equation}
\end{subequations}

\subsubsection{Distributions and Van Hove singularities}
\label{sec:vh1} 
\begin{figure}[htbp]
\begin{center}
\rotatebox{-90}{\includegraphics[width=7cm]{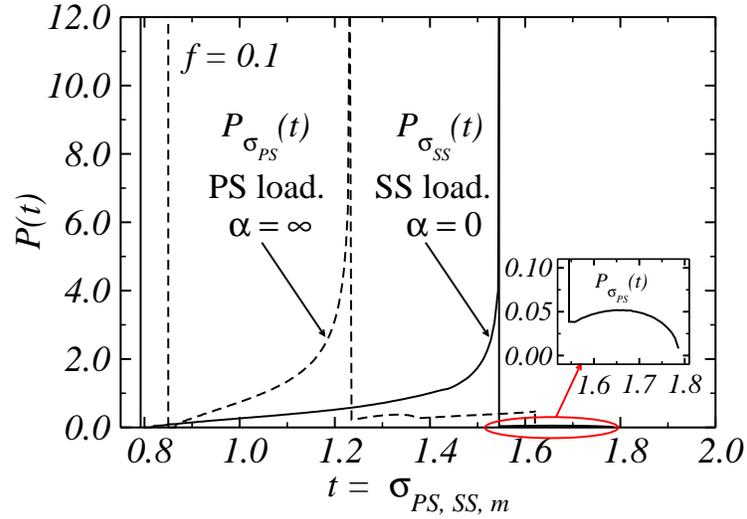}}
\begin{minipage}{12cm}
\caption{\label{fig:distributions_par} Distributions of the stress components
\emph{parallel} to the applied loading, at void concentration $f=0.1$, in
simple shear loading ($\langle\sigma_\text{SS}\rangle=1$)
 with $\alpha=0$ ($P_{\sigma_\text{SS}}$, solid), and in pure shear loading
 ($\langle\sigma_\text{PS}\rangle=1$)
 with $\alpha=\infty$ ($P_{\sigma_\text{PS}}$, dashed).
Straight vertical lines near $t=0.8$ indicate Dirac components. Inset:
magnification of the ``foot'' in $P_{\sigma_\text{SS}}(t)$.}
\end{minipage}
\end{center}
\end{figure}

\begin{figure}[htbp]
\begin{center}
\rotatebox{-90}{\includegraphics[width=7cm]{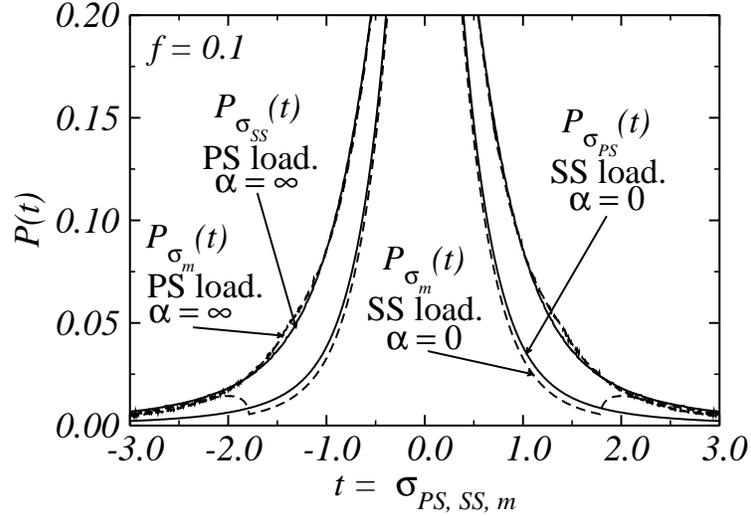}}
\begin{minipage}{12cm}
\caption{\label{fig:distributions_perp} Distributions of the stress
components \emph{transverse} to the applied loading, and of
$\sigma_\text{m}$, at void concentration $f=0.1$, in simple shear
loading ($\langle\sigma_\text{SS}\rangle=1$)
 with $\alpha=0$ ($P_{\sigma_\text{PS}}(t)\simeq P_{\sigma_\text{m}}(t)$
 for $|t|\lesssim 1.5$),
and in pure shear loading ($\langle\sigma_\text{PS}\rangle=1$) with
$\alpha=\infty$ ($P_{\sigma_\text{SS}}(t)\simeq P_{\sigma_\text{m}}(t)$). Dirac
components at $t=0$ should also be present in each of the four sets, due to the
void and to zone A, but are omitted for clarity.}
\end{minipage}
\end{center}
\end{figure}

The distributions of the stress fields in the matrix for SS
loading, at void concentration $f=0.1$, are displayed in Figs.\
\ref{fig:distributions_par} and \ref{fig:distributions_perp} as thick solid
lines.
In these figures, straight vertical lines represent Dirac components. The
distribution $P_{\sigma_\text{SS}}(t)$ of the stress ``parallel'' to the
applied loading (Fig.\ \ref{fig:distributions_par}) comprises one Dirac
component, proportional to $[(1-2a)^2/(1-f)]\delta(t-g_B(a))$, which represents
the contribution of zones (A) of constant $\sigma_\text{SS}$, and a divergence
followed by a discontinuity of infinite amplitude. The latter comes from the
(B) zones. The right ``foot'' (magnified in the inset) is produced by zones (D)
of maximal stress. The distribution has finite support since
$\varepsilon_\text{SS}$ is bounded. The distributions of the transverse
stresses $\sigma_\text{PS}$ and of $\sigma_\text{m}$ are displayed in Fig.\
\ref{fig:distributions_perp}. They are even (these fields average to 0). A
Dirac contribution at $\sigma=0$ (that of zone (A)) should be present in all
plots but is omitted for clarity. The main components blow up at $\sigma=0$ and
slowly decrease at $\sigma\to\infty$ (the contribution of zones (B),
essentially). The distribution $P_{\sigma_\text{m}}$ differs only slightly from
$P_{\sigma_\text{PS}}$ by the contribution of zones (D), the difference being
small at $f=0.1$, for the smallest values of the stress.

The above singular features are understood as follows. Tails of the
distributions at high values of the fields are determined by the square-root
singularities responsible for the infinite variances (\ref{eq:coneinfty}).
Definition (\ref{eq:distridef}) of the distribution generates Van Hove
singularities (VHS), in the vicinity of values $t=t_0$ for which there exists
extremal or saddle integration points $\mathbf{x}_i$ such that
$a(\mathbf{x}_i)=t_0$ and $\partial_\mathbf{x} a(\mathbf{x}_i)=0$. In 2D,
depending on the eigenvalues of the matrix $\partial^2_{\mathbf{x}\mathbf{x}}
a(\mathbf{x}_i)$ being of same or of opposite signs, the singularity in $P$ is
either a discontinuity or a logarithmic divergence (Van Hove, 1953).
``Extended'' Van Hove singularities (EVHS) (Abrikosov, Campuzano and Gofron,
1993) arise when in addition one or more eigenvalues of
$\partial^2_{\mathbf{x}\mathbf{x}} a(\mathbf{x}_i)$ vanish. Consider first the
strain distribution. Using (\ref{eq:exactGlobalMin1}), we obtain
\begin{equation}
g_B(z)=g_B(0)-\frac{\overline{\varepsilon}+g_B(a)}{a(1-2a)^2}z^2+O(z^4),
\end{equation}
so that in zone (B), for $a<x<1/2$, $0<y<a$,
\begin{equation}
\varepsilon_{xy}(x,y)\simeq
\frac{\overline{\varepsilon}+g_B(a)}{2(1-2a)}-\frac{\overline{\varepsilon}+g_B(a)}{2a(1-2a)^2}y^2+\ldots,
\end{equation}
of the generic type $ h(x,y)=h_0-h_1 y^2$, $h_1>0$. This generates an EVHS near
$t=h_0=(\overline{\varepsilon}+g_B(a))/[2(1-2a)]$, of the form
\begin{equation}
\label{eq:vh1} P_h(t)\simeq\int \frac{{\rm d}\!x\,{\rm
d}\!y}{V}\,\delta\bigl(h_0-h_1 y^2-t\bigr)=
\frac{L_x}{2V\sqrt{h_1}}(h_0-t)^{-1/2}\,\theta(h_0-t),
\end{equation}
where $L_x$ is the size of the integration domain on $x$ values. Only values
near $y=0$ contribute and the range of integration over $y$ need not be
completely specified. The four quadrants account for $L_x=2(1-2a)$, so that
\begin{equation}
P_{\varepsilon_{xy}}(t)\simeq
\frac{\sqrt{2a}(1-2a)^2}{(1-f)\sqrt{\overline{\varepsilon}+g_B(a)}}|t-h_0|^{-1/2}\theta(h_0-t).
\end{equation}
Hence the divergence of $P_{\sigma_\text{SS}}$, of square-root nature, and the
drop of infinite amplitude in Fig.\ \ref{fig:distributions_par} are captured by
this approximation, since
\begin{equation}
\label{eq:rescalP}
P_{\sigma_\text{SS}}(t)=P_{\varepsilon_{xy}}(t/(2\lambda))/(2\lambda).
\end{equation}
Consider next the field distributions $P_{\sigma_\text{PS}}(t)$ and
$P_{\sigma_\text{m}}(t)$ of Fig.\ \ref{fig:distributions_perp}, and examine the
contributions near $t=0$. According to (\ref{eq:sigmmpsB}) and to the remark
following, weak field values lie on the axes $x=0$, $y=0$, $x=\pm 1/2$, $y=\pm
1/2$, where the four points $(x,y)=(\pm 1/2,0)$ and $(x,y)=(0,\pm 1/2)$
contribute most. Focus for instance on $(x,y)=(1/2,0)$. From
(\ref{eq:sigmmpsB}), (\ref{eq:gBder}) and (\ref{eq:meanstressxy}), and for
$y\gtrsim 0$~:
\begin{equation}
\label{eq:expsmspsl=0} \sigma_\text{m}(x,y)=\sigma_\text{PS}(x,y)
\simeq\overline{\sigma}\frac{(1/2-x)y}{a(1-2a)^2}.
\end{equation}
This is a VHS of type $h(x,y)=h_1 x^2-h_2 y^2$ (rotated 45${}^o$)
with $h_1$, $h_2
> 0$ so that $P_h(t)\sim-\log|t|/(2V\sqrt{h_1 h_2})$ for $t\sim 0$, where
$V$ is the integration surface. Using (\ref{eq:expsmspsl=0}), the
contributions of the above four points gather into:
\begin{equation}
\label{eq:expsmspsl=0_2nd}
 P_{\sigma_\text{PS}}^{(B)}(t)\sim P^{(B)}_{\sigma_\text{m}}(t)\sim
-\frac{4a(1-2a)^2}{\overline{\sigma}
(1-f)}\log\left[2(1-2a)^2\frac{|t|}{\overline{\sigma}}\right],
\qquad t\to 0.
\end{equation}
The slowly-decaying tails at high $t$, of $P_{\sigma_\text{PS}}(t)$ and
$P_{\sigma_\text{m}}(t)$ in Fig.\ \ref{fig:distributions_perp} are due to
singularity (\ref{eq:singya}). Again focusing on the domain $a < x < 1/2$, $0<
y < a$ in zone (B), and expanding near $y=a$, we obtain from
(\ref{eq:sigmmpsB}) , (\ref{eq:gBder}) and (\ref{eq:meanstressxy})~:
$$
\sigma_\text{PS}(x,y)=\frac{\lambda}{4}(1-2x)g_B'(y)\sim - \frac{
\overline{\sigma}\sqrt{2a}}{4(1-2a)}\frac{1-2x}{\sqrt{a-y}}, \qquad
y\lesssim a.
$$
Of type $h(x,y)=-h_0(1-2x)/\sqrt{a-y}$, $h_0>0$, this expression
contributes for
\begin{equation}
\frac{1}{V}\int_a^{1/2}{\rm d}x \int^a_0{\rm d}y\, \delta\bigl(h(x,y)-t\bigr)
=\frac{h_0^2}{3V}\frac{(1-2a)^3}{|t|^3}\theta(-t)
\theta\left(|t|-\frac{(1-2a)h_0}{\sqrt{a}}\right),
\end{equation}
where the lower integration bound on $y$ provides the restriction on $|t|$
which delimits the domain of validity of the approximation. Adding
contributions from negative values, and multiplying by 4 due to square symmetry
leads to:
\begin{equation}
\label{eq:pb} P_{\sigma_\text{PS}}^{(B)}(t)\simeq
\overline{\sigma}^2\frac{a(1-2a)}{6(1-f)}|t|^{-3},\qquad
|t|\gg\frac{\overline{\sigma}}{2\sqrt{2}}.
\end{equation}
This is only the contribution of $(B)$. The tail should also include a
contribution from $(D)$ at high stresses, also decaying as $|t|^{-3}$, but
harder to handle due to sign changes at the points $(\pm a, \pm a)$. Other
singularities exist which have not been fully investigated: values of
$\sigma_\text{m}$ around $x,y=\pm a/\sqrt{2}$ produce a singularity in its
histogram (extra wings starting at $\sigma_\text{m}\simeq\pm 1.7$ with vertical
slope, see Fig.\ \ref{fig:distributions_perp}); moreover, for porosities $f$
higher than 0.6, two inflection points show up in $g_B(z)$. They give rise to
two local extrema of $\sigma_\text{m}$ and $\sigma_\text{PS}$ along the
cartesian axes zone (B), also producing extra wings of the above type (not
shown).

\subsection{Equibiaxial loading and limit $\lambda\to 0$}
\label{subsec:eqbxlam0}We close the study of the case $\alpha=0$ with
equibiaxial loading
$\overline{\varepsilon}=\langle\varepsilon_\text{m}\rangle\not=0$,
$\langle\varepsilon_{\rm PS}\rangle=\langle\varepsilon_{\rm SS}\rangle=0$. The
anisotropic limit $\alpha=0$ follows either from letting $\lambda\to 0$, or
$\mu\to\infty$, so that two different solutions may \textit{a priori} arise.
Only the case $\lambda\to 0$, with finite $\kappa$, is detailed in this
section. Case $\mu\to \infty$ is briefly addressed in Sec.\
\ref{subsec:eqbxmuinfty}. A method similar to that of Sec.\ \ref{sec:k0PS} for
PS loading is used. The differences relative to the PS case can be related to
the symmetries at play.

\subsubsection{Stress and strain fields}
The limit $\lambda=0$ under finite macroscopic mean stress requires that
$\sigma_{xy}\equiv 0$. Equilibrium and the equivalence of the $Ox$ and $Oy$
axes imply that:
\begin{equation}
 \sigma_{xx}(x,y)=g(y),\qquad \sigma_{yy}(x,y)=g(x),
\end{equation}
where $g(z)$ is defined on the interval $[-1/2,1/2]$. Similar steps
apply as in Sec.\ \ref{sec:k0PS}.
Energy minimization under the same
macroscopic stress constraint (\ref{eq:constrpureshear}) --- with
$\overline{\sigma}=\langle\sigma_\text{m}\rangle$ replacing
$\langle\sigma_{\rm PS}\rangle$, provides $\sigma_\text{SS}(x,y)=0$,
and
\begin{equation}
\label{eq:sigsol2}
 \sigma_\text{m,PS}(x,y)=\overline{\sigma}\,
\frac{\chi(y)\pm\chi(x)}{2(1-2a)},
\end{equation}
where the \textit{plus} (resp.\ \textit{minus}) sign applies to
$\sigma_\text{m}$ (resp.\ $\sigma_\text{PS}$). Associated strains follow from
(\ref{eq:lambdamu}). Periodicity conditions on $\mathsf{u}^*$ entail:
\begin{equation}
\label{eq:k0mem} \overline{\sigma}=\frac{2\mu(1-2a)}{m+(1-m)a}\overline{\varepsilon}.
\end{equation}
Eqn.\ (\ref{eq:sigsol2}) reproduces the PS result (\ref{eq:sigsol1}), with the
indices PS and m, and with $x$ and $y$, exchanged. Furthermore, the PS and
equibiaxial loading modes exchange the parts played by $\kappa$ and $\mu$,
which corresponds to interchanging $m\leftrightarrow 1/m$. This circumstance
almost allows for a one-to-one mapping of the results of this section onto
those of Sec.\ \ref{sec:k0PS}. For instance in the URQ:
\begin{equation}
\label{eq:ess} \varepsilon_\text{SS}=u_2
\left[\left(x-\frac{1}{2}\right)\delta(y-a)
+\left(y-\frac{1}{2}\right)\delta(x-a)\right],\quad
u_2=\frac{\overline{\varepsilon}}{4}\frac{1+m}{m+(1-m)a}.
\end{equation}
Eqn.\ (\ref{eq:ess}) differs from (\ref{eq:k0PSepspe}) by a substitution
$m\leftrightarrow 1/m$, and by a minus sign due to the difference between
(\ref{eq:a1g}) and (\ref{eq:a3g}), or between (\ref{eq:a1c}) and
(\ref{eq:a3c}).

\subsubsection{Displacement field and singularities}
The displacement $\mathbf{u}^\star$ in the URQ is such that $u_x(x,y)$ is there
of the form (\ref{eq:k0PSuxFull}), with $m$ replaced by $1/m$, but with now
$u^*_y(x,y)=u^*_x(y,x)$. Hence $\mathbf{u}^\star$ is again piecewise linear,
and tangentially discontinuous at the ``frontiers'' between (A), (B) and (D).
In particular, the jump $[[u^*_x]]_y(x,a)$ along the $y=a$ ``frontier'' in the
URQ is given by (\ref{jump}) with $m$ replaced by $1/m$. With $m=1$, the
overall pattern however differs from that of Fig.\ \ref{fig:taquin}, by the
fact that for $\overline{\varepsilon}<0$ (resp.\ $>0$), the displacements
$\mathbf{u}^\star$ in all parts of (B) are directed towards the void (resp.\
outwards), the medium being subjected to compression (resp.\ extension).

Moreover, owing to symmetry $u^*_y(x,y)=u^*_x(y,x)$, the PS mechanism
responsible for the formation of the square voids at $(x,y)=(\pm a,\pm a)$
changes into one which induces a compressive singularity near these locations,
\emph{independently of whether the equibiaxial loading mode is compressive or
extensive}, over a region of size $\Delta
u=\overline{\varepsilon}(m+1)(a/2-1/4)/[m+(1-m)a]$. This is readily seen from
the following expression valid in the vicinity of $(x,y)=(a,a)$:
\begin{equation}
\mathbf{u}(x,y)=u_2\left[ (a+1/2)\left(1,1\right)+(a-1/2)
 \left(\mathop{\text{sign}}(y-a),\mathop{\text{sign}}(x-a)
\right)\right].
\end{equation}
The displacement in (D) now reads $
\mathbf{u}(x,y)=u_2(\mathop{\text{sign}}x,\mathop{\text{sign}}y)$, so that the
four ``hot spots'' $(x, y)=(0, \pm a)$, $(\pm a, 0)$ now all undergo a similar
local singularity, either of a compressive, or of an extensive nature.

\subsubsection{Moments and effective compressibility modulus $\widetilde{\kappa}$}
The effective compressibility modulus and the moments are provided by
expressions (\ref{eq:PSmueff})-(\ref{eq:S1ss}), with $\mu$ and
$\widetilde{\mu}$ replaced by $\kappa$ and $\widetilde{\kappa}$, with $m$
replaced by $1/m$, and with the indices $PS$ and $m$ interchanged. For
definiteness:
\begin{equation}
\frac{\widetilde{\kappa}}{\mu}=\frac{(1-2a)}{m+(1-m)a}
=\frac{1}{m}\left[1-\frac{1+m}{m}(f/\pi)^{1/2}+O(f)\right].
\end{equation}
Thus, $\widetilde{\kappa}$ possesses a dilute-limit correction $\sim f^{1/2}$
at finite $\kappa$, but blows up as $f^{-1/2}$ in the incompressible limit
$m=0$. It decays as $(f_c-f)$ near $f_c$.

\subsection{Equibiaxial loading and limit $\mu\to\infty$}
\label{subsec:eqbxmuinfty} Consider now the limit $\mu\to\infty$, with finite
$\lambda$. A finite stress loading is consistent with this limit only if
$\varepsilon_\text{PS}\equiv 0$. This situation is that of $\alpha=0$ under SS
loading, the solution being of like complexity.In particular, for finite
compressibility, it is governed by elliptic equations (see Sec.\
\ref{sec:characteristics}) and is much more complicated than the one of Sec.\
\ref{subsec:eqbxlam0}. Accordingly, we restrict ourselves to the incompressible
case $\kappa=\infty$. The solution then simplifies and coincides with that of
Sec.\ \ref{subsec:eqbxlam0} with $m=0$, see (\ref{eq:lm}). Note that
$\widetilde{\kappa}$ is infinite, due to $\mu=\infty$.

\section{Material with anisotropy ratio $\alpha=\infty$}
\label{sec:ellinfty} This section is devoted to the case $\alpha=\infty$, where
the material is soft along the diagonals. As Fig.\ \ref{fig:bands}b
illustrates, the field patterns are now rotated by $45$ degrees. Accordingly,
the solutions obtained below could alternatively be derived in a frame where
the void lattice is rotated. The corresponding rotation, $\mathcal{R}_{45^{\rm
o}}$, would exchange in the matrix $\mathbb{E}^{\rm SS}$ and $\mathbb{E}^{\rm
PS}$, and $\lambda$ and $\mu$. In the rotated frame, the following
correspondence should then be used
\begin{equation}
\label{eq:symmetry45} \mathcal{R}_{45^{\rm o}}(\text{void lattice})
\Leftrightarrow \left\{ \alpha\leftrightarrow 1/\alpha,\text{ and } \text{PS
loading} \leftrightarrow \text{SS loading}, \right\}.
\end{equation}
Characteristics for the strain, stress or displacement fields are al\-ig\-ned
with the two diagonals of the unit cell (see Sec.\ \ref{sec:characteristics}).
Another difference with the case $\alpha=0$ resides in the appearance of new
regions in the matrix, denoted by (C) in the figure, where the (B) bands cross,
with no holes in it.

\subsection{Loading in simple shear}
We impose $\mu=0$, $\alpha=\infty$ and consider SS loading conditions. Due to
(\ref{eq:lambdamu}), $\mu=0$ implies
$\sigma_\text{PS}=(\sigma_{xx}-\sigma_{yy})/2\equiv0$, so that
$\sigma_{xx}(x,y)=\sigma_{yy}(x,y)\equiv s(x,y)$, an unknown function. Stress
equilibrium implies $\partial^2 s$ $/\partial x^2$ $-$ $\partial^2 s$
$/\partial y^2$ $=$ $0$. Using the identities in Appendix \ref{sec:sym}, we
deduce that $s(x,y)=[g(x-y)-g(x+y)]/2$, where $g$ is an unknown, even,
1-periodic function. The stress $\sigma_{xy}=\sigma_\text{SS}$ stems from
integrating the stress equilibrium equations. The integration constant is zero,
since $\mathbf{\sigma}=0$ in the void. Eventually, with $g(0)=0$,
\begin{equation}
\label{eq:siggalphainfty}
\sigma_\text{SS}(x,y)=[g(x-y)+g(x+y)]/2, \qquad
\sigma_\text{m}(x,y)=[g(x-y)-g(x+y)]/2.
\end{equation}
Due to the void one has, along the main diagonal,
$\sigma_\text{m}(x,x)=-g(2x)/2\equiv 0$ for $x<a/\sqrt{2}$. Hence $g(z)\equiv
0$ on $[-\sqrt{2}a,\sqrt{2}a]$. In turn, stresses (\ref{eq:siggalphainfty})
vanish in the whole square made of zones (V) and (D) in Fig.\
\ref{fig:bands}(b). Due to 1-periodicity and to
 $g$ being even, the stress pattern possesses a
mirror symmetry with respect to the dashed lines of Fig.\ \ref{fig:bands}(b).
Its fundamental unit cell is therefore the gray square delimited by them.
Symmetry (\ref{eq:symmetry45}) makes this case tantamount to the PS case of
Sec.\ \ref{sec:k0PS} with $\alpha=0$ and $\lambda=0$, with a cell size
diminished by a factor $1/\sqrt{2}$ at constant pore radius $a$ and,
consequently, with a porosity twice bigger. Effective moduli and moments are
then obtained from expressions (\ref{eq:momentsEll0PS}) of Sec.\
\ref{sec:k0PS}, with $\mu$ and $\widetilde{\mu}$ replaced by $\lambda$ and
$\widetilde{\lambda}$, with $m$ replaced by $\ell$, with the subscripts SS and
PS exchanged, and with $a$ replaced by $\sqrt{2}a$. Deformation in the
non-voided crossing zones (C) of zero stress is understood as follows. The
obtained solutions for the displacement and strain fields in the (D) zones are
trivially continued to the voided zone (V). These admissible continuations are
non-physical except on the void boundary. However, the above symmetry
considerations make clear that, modulo an appropriate translation, these
continuations also provide the physical deformation fields in (C).

\begin{figure}
\begin{center}
\epsfig{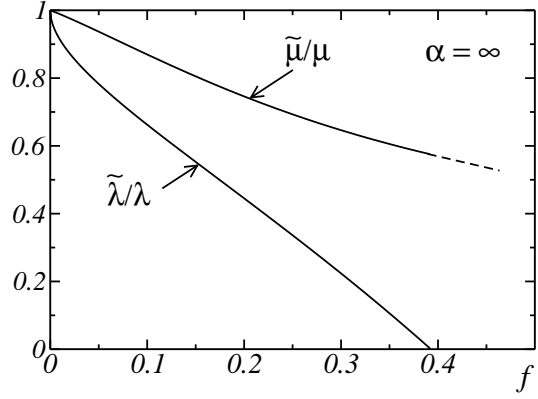}
\begin{minipage}{12cm}
\caption{\label{fig:effectiveModulivsfLinfty} Normalized shear
effective moduli $\widetilde{\mu}/\mu$ and
 $\widetilde{\lambda}/\lambda$ vs. void concentration
$f$, for an incompressible matrix with anisotropy ratio
$\alpha=\infty$. The curve for $\widetilde{\mu}/\mu$ is computed up
to $f=\pi/8$ only. Modulus $\widetilde{\lambda}$ vanishes at
$f_{c}^{(1)}= \pi/8\simeq 0.39$, \emph{half} the close packing
threshold value. }
\end{minipage}
\end{center}
\end{figure}

\textit{Mutatis mutandis}, the obtained expressions for the fields and the
moduli are valid only up to $f=f_c^{(1)}=\pi/8$, due to the rescaling of $a$.
This concentration acts as a ``mechanically driven percolation'' threshold, at
half the close-packing value. It corresponds to the configuration where the
summits (D) in Fig.\ \ref{fig:bands}b come into contact with the cell
boundaries. Then, (A) and (B) vanish, leaving only (C) and (D) for $f$ up to
$f_c=\pi/4$. Thus, for $f>f_c^{(1)}$, $\sigma_\text{SS}$ vanishes in the
matrix. The stress $\varepsilon_\text{SS}$ also vanishes, so that strain only
takes place along the void boundaries. Consequently, $\widetilde{\lambda}$
vanishes for $f\geq f_c^{(1)}$ and the fields are such that:
$M^{(1)}(\varepsilon_\text{SS})=0$, $M^{(1)}(\varepsilon_\text{PS})=1/f$,
$S^{(1)}(\varepsilon_\text{SS})=0$. Fig.\ \ref{fig:effectiveModulivsfLinfty}
displays $\widetilde{\lambda}(f)/\lambda$ vs.\ $f$. Except for the threshold,
the curve is alike that of $\widetilde{\mu}(f)/\mu$ in Fig.\
\ref{fig:effectiveModulivsfLzero}.

In words, we just showed that special field configurations in linear
anisotropic periodic media can lower the geometric ``percolation'' threshold of
the porous material (i.e.\ the close-packing threshold, for a periodic pore
lattice) into one determined by effective ``porous" zones (as far as they
undergo vanishing stresses) at intersections of void-generated stress bands.
This is unusual: for instance in homogenization methods, percolation-like
thresholds are usually thought (or found) to be independent of the constitutive
law.

\subsection{Loading in pure shear (incompressible case only)}
In the case $\alpha=\infty$ with $\lambda=\infty$ and $\mu$ finite, under PS
loading, the solution has the pattern of Fig.\ \ref{fig:bands}(b). It is of the
type studied in Sec.\ref{sec:k0SS}, at least for $f$ up to $f_c^{(1)}=\pi/8$.
For $f>f_c^{(1)}$ the problem becomes more involved due to a complex geometry,
and has not been investigated in full. Only the range $f<f_c^{(1)}$ is
discussed hereafter.

\subsubsection{Displacements, strain and stress fields}
The solution obeys $\varepsilon_\text{SS}=\varepsilon_{xy}\equiv 0$, and the
incompressibility constraint $\varepsilon_{xx}$ $+$ $\varepsilon_{yy}$ $\equiv$
$0$. Expressed in terms of $u^*_x$, $u^*_y$, such that $u^*_x(0,y)$ $=$
$u^*_y(x,0)$ $\equiv$ $ 0$ as implied by (\ref{eq:a1a}) and (\ref{eq:a1b}),
these equations entail with the help of (\ref{eq:a1c}):
\begin{equation}
u^*_x(x,y)=[G(x+y)+G(x-y)]/2,\quad u^*_y(x,y)=[-G(x+y)+G(x-y)]/2,
\end{equation}
and eventually with
$g(z)=G'(z)$ an even and 1-periodic function:
\begin{equation}
\label{eq:alphainftystrain}
\varepsilon_{\rm
PS}(x,y)=\left[g(y-x)+g(x+y)\right]/2.
\end{equation}
Introducing the 45${}^\textnormal{o}$ counterclockwise-rotated coordinates
$x'=(y+x)/\sqrt{2}$ and $y'=(y-x)/\sqrt{2}$, such that $-1/(2\sqrt{2})\leq
x',y'\leq 1/(2\sqrt{2})$, functions $g_\text{B}$ and $g_\text{A}$ are
introduced, such that
\begin{equation}
\label{eq:PSgagbdef}
g(\sqrt{2}x')=g_\text{B}(x')\,\theta_{[0,a]}(x')+g_\text{A}(x')\,\theta_{[a,1/(2\sqrt{2})]}(x').
\end{equation}
The total elastic energy in the unit cell (\ref{EffW}) is provided by the
integral:
\begin{equation}
(1-f)\langle w^{(1)}\rangle_{(1)}=2\mu
\int_M{\rm d}^2x\,\varepsilon_\text{PS}^2=2\mu
\biggl\{\qquad 2
\int_{\hspace{-3em}
[-1/(2\sqrt{2}),1/(2\sqrt{2})]^2\phantom{\biggl(}}\hspace{-4em}{\rm
d}x'{\rm d}y'-\int_V{\rm d}x'{\rm
d}y'\biggr\}\varepsilon_\text{PS}^2(x',y').
\end{equation}
One extra contribution of the void $V$ (of no elastic energy) has been
subtracted from the contribution of the gray square in Fig.\
\ref{fig:bands}(b), counted twice. Anticipating on the fact that on its
interval of definition $g_\text{A}(z)\equiv g_\text{B}(a)$, as in Sec.\
\ref{sec:k0SSdispl}, the energy (\ref{EffW}) is functionally minimized under
the constraint:
\begin{equation}
\label{eq:ebarPS}
\overline{\varepsilon}=\left\langle\varepsilon_\text{PS}\right\rangle=
2\sqrt{2}\int_0^a\hspace{-0.2cm}g_\text{B}(y){\,{\rm d}y}
+\left(1-2\sqrt{2}a\right)g_\text{B}(a),
\end{equation}
This average is computed on one gray square, with the strain field
(\ref{eq:alphainftystrain}) continued inside the void. Setting
$\rho(z)\equiv\sqrt{a^2-z^2}$, an integral equation results:
\begin{equation}
\label{eq:exactGlobalMin1PS} \sqrt{2}\int_0^{\rho(z)}\hspace{-0.3cm}
g_\text{B}(y){\,{\rm d}y}=\left[1-\sqrt{2}\rho(z)\right]
g_\text{B}(z)-g_\text{B}(a), \qquad z\in[0,a].
\end{equation}
Its solution is obtained following Sec.\ \ref{sec:k0SSdispl}. In
particular, with $z=0$ and (\ref{eq:ebarPS}):
\begin{equation}
\label{eq:gb0gBaPS}
g_\text{B}(0)=\frac{\overline{\varepsilon}
+(1+2\sqrt{2}a)g_\text{B}(a)}{2(1-\sqrt{2}a)}.
\end{equation}
The displacement is everywhere continuous. The stress $\sigma_\text{PS}$ is
deduced from (\ref{eq:alphainftystrain}) and from
$\sigma_\text{PS}=2\mu\varepsilon_\text{PS}$ in the matrix, with
$\sigma_\text{PS}=0$ in the voids. The transverse components $\sigma_\text{m}$,
$\sigma_\text{SS}$ are computed analogously to (\ref{eq:sigmak0SS}), with
integrations and differentiations carried out over $x'$ and $y'$. Stress
equilibrium provides:
\begin{equation}
\partial_{x'}\,\sigma_\text{PS}=  \partial_{y'}(\sigma_\text{m}-\sigma_\text{SS}), \qquad
\partial_{y'}\,\sigma_\text{PS} =
\partial_{x'}(\sigma_\text{m}+\sigma_\text{SS}).
\end{equation}
Due to symmetry, $\sigma_\text{m}(x,x)=\sigma_\text{SS}(x,x)\equiv
0$ and $\sigma_\text{m}(x,1-x)=\sigma_\text{SS}(x,1-x)\equiv 0$, so
that with  $0<x,y<1$ in the matrix (for convenience, a unit cell
translated by a vector $\mathbf{t}=(1/2,1/2)$ is considered):
\begin{equation}
\sigma_\text{m,SS}(x',y')=(1/2)
 \int_{1/\sqrt{2}}^{x'}\!\!\!{\rm d}z\,\partial_{y'}\sigma_\text{PS}(z,y')
 \pm(1/2)\int_0^{y'}\!\!\!{\rm
 d}z\,\partial_{x'}\sigma_\text{PS}(x',z),
\end{equation}
The \emph{plus} (resp.\ \emph{minus}) sign applies to $\sigma_m$
(resp.\ $\sigma_\text{SS}$). Then, in the matrix:
\begin{equation} \label{eq:SssSm}
\sigma_\text{m,SS}(x,y)=\frac{\mu}{2} \left[
(x+y-1)g'\left(y-x\right)\pm(y-x)g'\left(x+y-1\right)
 \right]
\end{equation}
for $0<x,y<1$. For $z$ in $[0, a\sqrt{2}]$, the derivative $g'$ reduces to $
g'(z)$ $=$ $g'_\text{B}(z/\sqrt{2})$ $/\sqrt{2}$ where $g'_B$ is the derivative
of $g_B$. The stress and strain singularities are qualitatively the same as in
the $\alpha=0$, SS loading case. In particular, from
(\ref{eq:exactGlobalMin1PS}), $g_B(z)\simeq g_B(a)+2\sqrt{a}[g_B(0)+
g_B(a)]|a-z|^{1/2}$, for $z\lesssim a$, so that $\sigma_\text{PS}$ blows up
along the ``frontiers'' of (A), (B), (C) and (D) as $d^{-1/2}$ where $d$ is the
distance to the ``frontier''.

\subsubsection{Moments and effective modulus $\widetilde{\mu}(f)$ in the dilute limit}
Expanding $g_B(z)$ in powers of $a$, we obtain:
\begin{equation}
\label{eq:epsBaPS}g_B(a)/\overline{\varepsilon}=1-2\pi
a^2-\frac{32\sqrt{2}}{3}a^3+
 \left[(5/2)\pi^2-6\pi-8\right]a^4+O(a^5).
\end{equation}
Dilute expansions in $f$ for the effective modulus and the moments
ensue:
\begin{subequations}
\begin{equation}
\frac{\widetilde{\mu}}{\mu}=1-f-\frac{32}{3\sqrt{2}\pi^{3/2}}f^{3/2}+
\left(1-\frac{3}{\pi}-\frac{4}{\pi^2}\right)f^2+O(f^{5/2}),
\end{equation}
\begin{equation}
M^{(1)}(\varepsilon_\text{PS})=1-\frac{32
f^{3/2}}{3\sqrt{2}\pi^{3/2}}+ O(f^2),\quad
M^{(2)}(\varepsilon_\text{PS})=1+\frac{32
f^{1/2}}{3\sqrt{2}\pi^{3/2}}+ O(f),
\end{equation}
\begin{equation}
S^{(1)}(\varepsilon_\text{PS})= \frac{4\, 2^{1/4}
f^{3/4}}{\sqrt{3}\pi^{3/4}}+ O(f^{5/4}),\quad
S^{(1)}(\sigma_\text{PS})=\frac{4\,
2^{1/4}f^{3/4}}{\sqrt{3}\pi^{3/4}}+O(f^{5/4}),
\end{equation}
\begin{equation}
S^{(1)}(\varepsilon_\text{SS})=0,\quad
S^{(1)}(\sigma_\text{SS})=\infty; \quad
S^{(1)}(\varepsilon_\text{m})=0,\quad
S^{(1)}(\sigma_\text{m})=\infty.
\end{equation}
\end{subequations}
In the range $0<f<\pi/8$, $\widetilde{\mu}(f)$ is computed from a numerical
solution of (\ref{eq:exactGlobalMin1PS}). Contrarily to the SS case where
$\widetilde{\lambda}(f)$ vanishes at $\pi/4$, no \ ``mechanically driven
percolation'' is found here, since the zones (C) where bands cross are not
zones of vanishing stress. Regular close packing behavior must occur at
$f_c=\pi/4$.

\subsubsection{Distributions}
\label{sec:vh2} Field distributions are studied as in Sec.\ \ref{sec:vh1}.
Quite similar features are observed, summarized hereafter, along with some
differences. The density $P_{\varepsilon_\text{PS}}$ is non-symmetric, and
supported by the interval $[g_\text{B}(a),g_\text{B}(0)]$. A rescaling similar
to (\ref{eq:rescalP}), using $\mu$ instead of $\lambda$, provides
$P_{\sigma_\text{PS}}$, plotted in Fig.\ \ref{fig:distributions_par} (dashed
curve). Due to the gradient of $\varepsilon_\text{PS}(x,y)$ vanishing along the
lines $y=\pm x$, where the field is maximum in (B),
$P_{\varepsilon_\text{PS}}(t)$ blows up as $|t-t_0|^{-1/2}$ at
$t_0=[g_\text{B}(a)+g_\text{B}(0)]/2$. A notable difference with the case
$\alpha=0$ lies in the existence of the extended ``foot'' on the right side of
$P_{\sigma_\text{PS}}$. It ends with a discontinuous VHS of finite amplitude,
due to the maximum strain field value,
$\max\varepsilon_\text{PS}=g_\text{B}(0)$, being reached at $(x,y)=(\pm 1/2,\pm
1/2)$, in (C). In the case $\alpha=0$, $\sigma=0$ in this zone. The
discontinuity is computed by remarking that according to
(\ref{eq:alphainftystrain}), (\ref{eq:PSgagbdef}) and to a Taylor expansion of
$g_B(z)$ at $z=0$, the strain field $\varepsilon_\text{PS}$ at points
$(x,y)=(\pm 1/2,\pm 1/2)$ is a quadratic function with negative eigenvalues.
Furthermore, a field $h(x,y)\simeq-(h_1x^2+h_2y^2)$ with $h_1$, $h_2>0$, on a
domain $V$, produces in $P_h(t)$ a jump at the origin
$[[P_h(0)]]=-\pi/(4\sqrt{h_1h_2}V)$.

Densities $P_{\sigma_\text{m}}$, $P_{\sigma_\text{SS}}$ stem from
Eqn.\ (\ref{eq:SssSm}). They blow up at the origin due to points of
 vanishing derivatives $\partial_x \sigma_\text{m,SS}$, $\partial_y
\sigma_\text{m,SS}$. According to (\ref{eq:SssSm}) this happens for
$(x,y)=(1/2,1/2)$. In the vicinity of this point, the expansion of
$\sigma_\text{m}$ is quadratic of a regular type and leads to logarithmic
divergent VHS of $P_{\sigma_\text{m}}(t)$ at the origin. On the other hand,
$\sigma_\text{SS}(x,y)$ is of an unusual cubic form:
\begin{equation}
\sigma_\text{SS}(x,y)\propto (y-x)(x+y-1)\left(x-1/2\right).
\end{equation}
This generates a singular contribution to $P_{\sigma_\text{SS}}(t)$ at the
origin, of a type not previously encountered: indeed, the distribution $P_h(t)$
of a function $h(x,y)=h_0(y^2-x^2)x$ behaves near $t=0$ as $|t|^{-1/3}$.
Finally, the tails of $P_{\sigma_\text{m}}(t)$, $P_{\sigma_\text{SS}}(t)$ when
$|t|\to\infty$ decay unsurprisingly as $|t|^{-3}$, due to the blowing-up of the
concerned stress components along the ``frontiers'' between zones (A), (B),
(C), (D) as an inverse square root of the distance.

\subsection{Equibiaxial loading}
We consider here the limit $\alpha=\lambda/\mu=\infty$, achieved by taking
$\mu\to 0$, under equibiaxial loading. The limit $\lambda\to \infty$ is not
dealt with for the same reason as in Sec.\ \ref{subsec:eqbxmuinfty}. Moderate
porosities $f<f_c$ are assumed. The method used for the case $\alpha=0$ is
applied to the void lattice rotated by 45 degrees, and undergoing symmetry
(\ref{eq:symmetry45}). With $a_s=\sqrt{2}a$ and with the rotated coordinates
$x'=(y+x)/\sqrt{2}$ and $y'=(y-x)/\sqrt{2}$, the stresses are
$\sigma_\text{PS}(x,y)\equiv 0$, and
\begin{equation}
\sigma_\text{m,SS}(x,y)=\overline{\sigma}[\chi(y')\pm\chi(x')]
/[2(1-2a_s)],
\end{equation}
in the rotated gray domain of Fig.\ \ref{fig:bands}b. Hence, with
$\varpi\equiv (\overline{\varepsilon}/2)/[\ell+(1-\ell)a_s]$,
\begin{equation}
\varepsilon_\text{m}(x,y)=\ell\,\varpi\,[\chi(y')+\chi(x')],\quad
\varepsilon_\text{SS}(x,y)=\varpi\,[\chi(y')-\chi(x')].
\end{equation}
Likewise, the PS component of the strain field takes the Dirac-localized form
for $x',y'$ in the URQ of the rotated gray square:
\begin{equation}
\varepsilon_\text{PS}(x,y)=(1+\ell)\varpi
\left[\left(x'-\frac{1}{\sqrt{2}}\right)\delta(y'+a)
+\left(y'-\frac{1}{\sqrt{2}}\right)\delta(x'-a)\right].
\end{equation}
Then, the following effective compressibility modulus is obtained:
\begin{equation}
\frac{\widetilde{\kappa}}{\lambda}=\frac{1-2a_s}{\ell+(1-\ell)a_s}.
\end{equation}
``Mechanically driven percolation'' is again observed here. The moments of the
fields are read from expressions (\ref{eq:PSfieldmeans})--(\ref{eq:S1ss}), by
carrying out the following substitutions: (i) replace $a$ by $a_s$; (ii)
replace $m$ by $1/\ell$; (iii) replace index PS by index m; (iv) replace index
m by index SS; (v) replace index SS by index PS.

\section{Discussion and conclusion}
\label{sec:concl} Two essentially different types of solutions, in an
infinitely anisotropic linear elastic medium containing a periodic distribution
of voids, have been exhibited. The anisotropy directions of the matrix --- a
``hard'' one and a ``soft'' one --- being aligned with the lattice directions,
the fields are arranged in patterns of bands.

On the one hand, for shear loading along a ``hard'' direction, the following
observations were made: ($i$) the ``parallel'' components of the stress and
strain are discontinuous, piecewise constant, with a band width of one pore
diameter; ($ii$) the deformation pattern has a ``block sliding'' structure;
($iii$) the shear strain orthogonal to the loading has infinite variance, being
localized as Dirac distributions along the band ``frontiers'', which are
tangent to the voids (with jumps at the tangency points); ($iv$) the
``perpendicular'' component of the stress is zero. The presence of
discontinuities in the tangential component of the displacement field is
analogous to similar observation in ideal plasticity, including the Hencky
plasticity model (Suquet, 1981). Moreover, in situations where the bands cross,
the medium develops fictitious porous zones of zero parallel stress which lower
the close packing threshold. Also, a first-order correction $\sim f^{1/2}$ is
induced in the effective shear modulus. The latter decays linearly with the
void concentration near the close packing threshold.

On the other hand, for loading along a ``soft'' direction, another type of
solution is produced when the problem is hyperbolic, i.e. assumed to be
incompressible. In this case, the following observations were made: ($i$) the
strains are continuous everywhere; ($ii$) the leading order correction to the
effective shear modulus in the loading direction is now $\propto f$, i.e.\ less
sensitive to increasing void concentration for small concentrations; ($iii$)
the variances of the ``parallel'' stress and strain in the matrix are $\propto
f^{3/4}$ and are again less sensitive to increasing void concentration; ($iv$)
the parallel strain enjoys scaling properties near the close-packing threshold
$f_c$, operating within a band of width $\xi\sim a(f_c-f)^{1/2}$, and becomes
fully localized only for $f$ near $f_c$. This situation is at best one of
``very soft'' void-induced \emph{strain} localization. However, stresses
accumulate in the ``hard'' directions (transverse and equibiaxial) and blow up
along the ``frontiers'' of the field pattern as the inverse square root of the
distance (the corresponding strains being zero). This particular \emph{stress}
localization configuration may be relevant to strain-locking materials, such as
the shape-memory polycrystals that have been considered recently by
Bhattacharya and Suquet (2005) and Chenchiah and Bhattacharya (2005).

In conclusion, situations responsible for fractional exponents showing up in
a\-ni\-so\-tro\-pic linear theories, of relevance to non-linear homogenization
methods, have been clarified. Comparisons with fully numerical results at
moderate anisotropies, and for isotropic viscoplasticity, will be presented
elsewhere (Willot et al., 2007). As a final remark, the fact that strong
singularities are found (isolated points of matter overlapping or of matter
separation) in one case is obviously of relevance for breakdown studies (e.g.,
in brittle materials), since these points may act as initiators. This may
suggest that mechanically sounder solutions could be looked for in a large
deformation framework. However, in view of our more modest aim to help better
understand possible hallmarks of localization in non-linear homogenization
theories, the model considered here is adequate.

\section{Acknowledgements}
The authors wish to thank the anonymous referee for remarks pertaining to the
interpretation of the solutions in terms of characteristics. The work of
P.P.C.\ was supported in part by the N.S.F.\ through Grant OISE-02-31867 in the
context of collaborative project with the C.N.R.S.


\appendix
\section{Symmetry properties for the strain, stress and displacement fields}
\label{sec:sym} Since the periodic anisotropic medium obeys the symmetries of
the square, the symmetries of the displacement $\mathbf{u}$ are determined by
the applied loading. They are easily deduced from Figs.\
\ref{fig:microstructure}(b) in simple shear and 1(c) in pure shear, and from an
obvious flow pattern in equibiaxial loading. Differentiations wrt.\ $x$ and $y$
provides those enjoyed by $\varepsilon_{ij}$. The symmetry group of the
constitutive law carries them unchanged over $\sigma_{ij}$. For
$\mathbf{v}=\mathbf{u}$ or $\mathbf{u}^*$ and for $\mathsf{a}=\mbox{\boldmath
$\varepsilon$}$ or $\mbox{\boldmath $\sigma$}$:
\begin{itemize}
\item \emph{In PS loading}:
\begin{subequations}
\label{eq:sym3}
\begin{eqnarray}
\label{eq:a1a}
v_x(x,y)&=&-v_x(-x,y)=v_x(x,-y), \\
\label{eq:a1b}
v_y(x,y)&=&v_y(-x,y)=-v_y(x,-y), \\
\label{eq:a1c}
v_x(x,y)&=&-v_y(y,x),\\
a_{ii}(x,y)&=&a_{ii}(-x,y)=a_{ii}(x,-y),\qquad i=x,y,\\
\label{eq:a1e}
a_{xx}(x,y)&=&-a_{yy}(y,x),\\
\label{eq:a1f}
a_{xy}(x,y)&=&-a_{xy}(-x,y)=-a_{xy}(x,-y),\\
\label{eq:a1g}
a_{xy}(x,y)&=&-a_{xy}(y,x);
\end{eqnarray}
\end{subequations}
\item \emph{In SS loading}:
\begin{subequations}
\begin{eqnarray}
\label{eq:a2a}
v_x(x,y)&=&v_x(-x,y)=-v_x(x,-y),\\
\label{eq:a2b}
v_y(x,y)&=&-v_y(-x,y)=v_y(x,-y),\\
\label{eq:a2c}
v_x(x,y)&=&v_y(y,x),\\
\label{eq:a2d}
a_{ii}(x,y)&=&-a_{ii}(-x,y)=-a_{ii}(x,-y),\qquad i=x,y,\\
\label{eq:a2e}
a_{xx}(x,y)&=&a_{yy}(y,x),\\
\label{eq:a2f}
a_{xy}(x,y)&=&a_{xy}(-x,y)=a_{xy}(x,-y),\\
\label{eq:a2g}
a_{xy}(x,y)&=&a_{xy}(y,x);
\end{eqnarray}
\end{subequations}
\item \emph{In equibiaxial loading}:
\begin{subequations}
\begin{eqnarray}
\label{eq:a3a}
v_x(x,y)&=&-v_x(-x,y)=v_x(x,-y),\\
\label{eq:a3b}
v_y(x,y)&=&v_y(-x,y)=-v_y(x,-y),\\
\label{eq:a3c}
v_x(x,y)&=&v_y(y,x),\\
\label{eq:a3d}
a_{ii}(x,y)&=&a_{ii}(-x,y)=a_{ii}(x,-y),\qquad i=x,y,\\
\label{eq:a3e}
a_{xx}(x,y)&=&a_{yy}(y,x),\\
\label{eq:a3f}
a_{xy}(x,y)&=&-a_{xy}(-x,y)=-a_{xy}(x,-y),\\
\label{eq:a3g}
a_{xy}(x,y)&=&a_{xy}(y,x).
\end{eqnarray}
\end{subequations}
\end{itemize}
\vfill \eject


\begin{thebibliography}{99}
\bibitem{GOFR93} Abrikosov, A.A., Campuzano, J.C., Gofron, K., 1993.
Experimentally observed extended saddle point singularity in the
energy spectrum of YBa${}_2$Cu${}_4$O${}_{6.9}$ and
YBa${}_2$Cu${}_4$O${}_{8}$ and some of the consequences, Physica C:
Superconductivity 214 (1-2), 73--79.

\bibitem{BHATTSUQ05} Bhattacharya, K. and Suquet, P. M., 2005.
A model problem concerning recoverable strains of shape memory
polycrystals.
Proc. R. Soc. Lond. A 461, 2797--2816.

\bibitem{CHEN05}
Chenchiah, I.V. and Bhattacharya, K., 2005. Examples of non-linear
homogenization involving degenerate energies. I. Plane strain. Proc.\ R.\ Soc.\
A 461, 3681--3703.

\bibitem{CULE98}
Cule, D., Torquato, S., 1998. Electric field distribution in
composite media. Phys.\ Rev.\ B 58, 11829--11832.

\bibitem{DAY92} Day, A.R., Snyder K.A., Garboczi E.J., Thorpe M.F., 1992.
The elastic moduli of a sheet containing circular holes.
J.\ Mech.\ Phys.\ Solids 40, 1031--1051.

\bibitem{DRUC66} Drucker, D.C., 1966.
The continuuum theory of plasticity on the macroscale and the microscale.
J.\ Mater.\ 1, 873--910.

\bibitem{EKE74} Ekeland, I. and Temam, R., 1974.
\emph{Analyse convexe et probl\`emes variationnels}.
Dunod, Paris.

\bibitem{FRAN98}
Francescato, P., Pastor, J., 1998. R\'esistance de plaques multiperfor\'ees~:
comparaison calcul-experience. Rev.\ Eur.\ \'El\'ements Finis 7, 421--437.

\bibitem{FRAN04} Francescato, P., Pastor, J., Riveill-Reydet, B., 2004.
Ductile failure of cylindrically porous materials. Part I: plane stress problem
and experimental results. Eur.\ J.\ Mech.\ A/Solids 23, 181--190.

\bibitem{KAC74} Kachanov, L.M., 1974.
\emph{Fundamentals of the theory of plasticity}.
Mir Publishers, Moscow.

\bibitem{LAHE332}
Lahellec, N., Suquet, P., 2004.
Nonlinear composites: a linearization procedure
exact to second order in contrast and for which the strain-energy and affine
formulations coincide. C.R.\ M\'ecanique 332, 693--700.

\bibitem{MACP94}
McPhedran, R.C. and Movchan, A.B., 1994. The Rayleigh multipole method for
linear elasticity. J. Mech. Phys. Solids 42, 71 l--727.

\bibitem{MASS00}
Masson, R., Bornert, M., Suquet, P., Zaoui, A., 2000. An affine formulation for
the prediction of the effective properties of nonlinear composites and
polycrystals. J.\ Mech.\ Phys.\ Solids.\ 48, 1203--1227.

\bibitem{OTTO03} M. Otto, J.-P. Bouchaud, P. Claudin, J. E. S. Socolar, 2003.
Anisotropy in granular media: Classical elasticity and directed-force chain
network. Phys.\ Rev.\ E 67, 031302.

\bibitem{PAST02}
Pastor, J., Ponte Casta\~neda, P., 2002. Yield criteria for porous media in
plane strain: second-order estimates versus numerical results. C.R.\
M\'ecanique 330, 741--747.

\bibitem{PELL01} Pellegrini, Y.P., 2001.
Self-consistent effective-medium approximation for strongly non-linear media.
Phys.\ Rev.\ B 64, 134211.

\bibitem{PONT91} Ponte Casta\~neda, P., 1991.
The effective mechanical properties of nonlinear isotropic composites. J.\
Mech.\ Phys.\ Solids.\ 39 (1), 45--71.

\bibitem{PONT92} Ponte Casta\~neda, P., DeBotton, G., Li, G., 1992.
Effective properties on nonlinear inhomogeneous dielectrics. Phys.\ Rev.\ B 46
(8), 4387--4394.

\bibitem{PONT96} Ponte Casta\~neda, P., 1996.
Exact second-order estimates for the effective mechanical properties of
nonlinear composite materials. J.\ Mech.\ Phys.\ Solids 44, 827--862.

\bibitem{PONT01} Ponte Casta\~neda, P., 2001.
Second-order theory for nonlinear dielectric composites incorporating field
fluctuations. Phys.\ Rev.\ B 64, 214205.

\bibitem{PONT02} Ponte Casta\~neda, P., 2002.
Second-order homogenization estimates for nonlinear composites incorporating field
fluctuations. I Theory. J.\ Mech.\ Phys.\ Solids 50, 737--757, and Part II
-Applications, ibid., 759--782.

\bibitem{SUQ81} Suquet, P.M., 1981.
Sur les \'equations de la plasticit\'e: existence et r\'egularit\'e des
solutions. J.\ M\'ecanique\ 20, 3--39.

\bibitem{TORQ02} Torquato, S., 2002.
\emph{Random Heterogeneous Materials: microstructure and macroscopic properties}.
Springer, New York.

\bibitem{VANH53} Van Hove, L., 1953.
The occurrence of singularities in the elastic frequency distribution of a
crystal. Phys.\ Rev.\ 89 (6), 1189--1193.

\bibitem{WILL86}
Willis, J. R., 1986. Variational estimates for the overall response of an
inhomogeneous nonlinear dielectric. In: \emph{Homogenization and Effective
Moduli of Materials and Media}, J.L. Ericksen {\it et al} eds.,
 Springer-Verlag, New-York, 247--263.

\bibitem{WILL07}
Willot, F., Pellegrini, Y.-P., Idiart, M., Ponte Casta\~neda,
P., 2007. In preparation.

\bibitem{ZACH86}
Zachmanoglou, E. C., Thoe, D.W., 1986. \emph{Introduction to partial
differential equations}. Dover, New York.

\bibitem{ZENG88} Zeng, X.C.,Bergman, D.J., Hui, P.M., Stroud, D., 1988.
Effective-medium theory for weakly non-linear composites.
Phys.\ Rev.\ B 38, 10970--10973.
\end{thebibliography}
\end{document}